\documentclass[11pt]{article}
\usepackage{arxiv}
\usepackage[utf8]{inputenc} 
\usepackage[T1]{fontenc}    
\usepackage{hyperref}       
\usepackage{url}            
\usepackage{booktabs}       
\usepackage{amsfonts}       
\usepackage{nicefrac}       
\usepackage{microtype}      
\usepackage{lipsum}
\usepackage{amsmath}
\usepackage[dvipsnames]{xcolor}
\usepackage{tikz}
\usepackage{subcaption}
\usepackage{float}
\usepackage{amssymb}
\usepackage{graphicx}
\usepackage{calc}
\usepackage{subcaption}
\usepackage{graphicx}
\usepackage[outdir=./]{epstopdf, epsfig}
\usepackage{amsmath}
\usepackage[dvipsnames]{xcolor}
\usepackage{tikz}
\usepackage{subcaption}
\usepackage{float}
\usepackage{todonotes}
\usepackage{lipsum}
\usepackage[export]{adjustbox}

\title{Modelling two-dimensional droplet rebound off deep fluid baths}
\author{
K. A. Phillips$^{1,2,\dagger}$, R. Cimpeanu$^{1,3,4}$  and P. A. Milewski$^{5}$ \\
\And 
$^\dagger$corresponding author : kat.phillips@warwick.ac.uk \\
$^{1}$Mathematics Institute, University of Warwick, Coventry CV4 7AL, United Kingdom\\
$^{2}$ Department of Mathematical Sciences, University of Bath, Claverton Down, Bath, BA2 7AY \\
$^{3}$Mathematical Institute, University of Oxford, Oxford OX2 6GG, United Kingdom\\
$^{4}$Department of Mathematics, Imperial College London, London SW7 2AZ, United Kingdom\\
$^{5}$Department of Mathematics, The Pennsylvania State University, State College, PA 16802
USA
}
\date{}

\begin{document}
\maketitle
\begin{abstract}
In order for a droplet to rebound rather than coalesce with a liquid bath, a layer of gas must persist throughout the impact. This gas, typically an air layer acts as a lubricant to the system and permits a pressure transfer between the two liquid bodies. Through considering separately the bath, air, and drop regions of fluid, we introduce a fully coupled reduced dynamic model of two-dimensional droplets (i.e. cylindrical geometry) rebounding off liquid baths, which incorporates an evolving lubricating air layer. Numerical solutions of the lubrication-mediated model are compared to dedicated direct numerical simulation of the Navier-Stokes equations. The reduced model captures rebounding dynamics well in the regime where it is most relevant: for low-speed impacts of small droplets, where capillary forces are important. Numerically, the reduced model is efficient, allowing for the computation of multiple rebounds and of long time dynamics of droplets rebounding on a vibrating bath. Furthermore, the lubrication-mediated model is able to provide detailed information within the air layer such as pressure and lubrication-layer geometry, which is usually omitted from reduced models.
\end{abstract}

\section{Background and motivation}\label{sect:Introduction}


Since the work of Worthington \cite{worthington1881impact, worthington1883impact}, we have endeavoured to classify the different behaviours that arise when droplets impact a liquid surface. An important question is to distinguish under which conditions a droplet will rebound versus coalesce or splash \cite{rein1993phenomena}, especially in industrial processes such as spray injections \cite{moreira2010advances}.

The ability for a droplet to rebound depends entirely on whether the layer of air that separates the two fluids remains unbroken throughout the impact. In the event that this air layer becomes sufficiently thin, the Van der Waals forces between the two liquid regions act, and the liquid pierces through the air layer and initiates coalescence \cite{sprittles2024gas}. For rebound to occur, it is a delicate balance between the system having enough restoring force to jettison the drop upwards after a cushioned impact, but not so much that the air layer is breached due to the inertia of the drop before the liquid surface has time to force the droplet upwards. 
"inviscid fluid model for the water entry of a rigid, solid body into a quiescent pool

Industrial interest in solid impact dynamics were initially heavily motivated by military interest in sea-plane landings and slamming of boat hulls, leading to the development of Wagner theory \cite{wagner1932phenomena, korobkin2004analytical} as a foundational inviscid model for solid body entry into quiescent liquids. The role of the air layer in pre-impact dynamics of solid bodies on deep liquid baths was first developed by Smith et al. \cite{smith2003air}. For this work we base ourselves on the investigation of Moore \cite{moore2021introducing}, who combined pre-impact cushioning with the Wagner model. We incorporate surface tension effects in drop oscillations and in wave generation and propagation in the bath, relevant to scenarios such as spray coating and ink-jet printing (see Derby\cite{derby2010inkjet}) where surface tension dominates small droplet behaviour.. Additionally, studies of droplet-droplet impacts have been motivated by the rich field of aerosols in a wide variety of natural processes from cloud seeding \cite{bird2010daughter} to pathogen transfer through suspended droplets \cite{yang2007size}.

This study is originally motivated by phenomena where a vertical vibration of the liquid bath permits sustained repeated droplet bouncing to occur \cite{couder2005walking}. These Faraday pilot-wave systems, named due to the Faraday waves generated at sufficiently strong vibration, can present a wide range of complex behaviours. To obtain a strongly coupled drop-bath system, the vibration is held below that which destabilises the free surface (the Faraday threshold). Within this regime, there are multiple drop-bath behaviours that can be experimentally achieved and modelled, and have been quantified in work such as those of Protiere et al. \cite{protiere2006particle}, and Milewski et al. \cite{milewski2015faraday}. As the forcing reaches near the Faraday threshold, the bouncing modes are destabilised, and through the asymmetry of impact a horizontal force is applied to the drop, permitting them to `walk' along the surface of the bath, guided by the waves generated from previous bounces. This Faraday pilot-wave system has drawn parallels to the quantum theory described by de Broglie \cite{de1926interference}. Dubbed the hydrodynamic quantum analogue, attempts have been made to recreate macro-scale analogies to quantum phenomena. For comprehensive reviews see the works of Bush \cite{bush2015pilot, bush2018introduction} and references therein.
The nature of these studies has required long-time simulations of repeated bouncing, which quickly become computationally expensive. Efforts to minimise this cost include reducing the number of spatial dimensions, such as laid out in the work by Nachbin et al. \cite{nachbin2017tunneling}, where a two-dimensional system was used to investigate the relationship between a droplet and the bath topography and the resultant tunnelling effects between bath cavities, or linking coupled droplet dynamics \cite{nachbin2018walking}, and recent works have highlighted the importance of including vertical dynamics within these methods \cite{papatryfonos2024static}. 

Two-dimensional configurations provide comparatively fast, detailed understanding of the key behaviours which are computationally expensive in three-dimensions. This approach can also be viewed as the  `infinite-cylinder' approximation of the geometry as a two dimensional droplet represents the cross-section of a cylinder. This consideration in itself can be well-motivated by physical phenomena such as water-walkers, whose long feet are cylindrical to leading order \cite{hu2003hydrodynamics}.

The linear inviscid model for the oscillations of a droplet was developed in 1924 by Lamb\cite{lamb1924hydrodynamics}. The model has been validated experimentally by work such as in Becker et al. \cite{becker1991experimental}. Considering droplets in zero gravity cases allows for the uninterrupted study of their oscillatory nature \cite{wang1996oscillations} and the longer viscous damping timescales. Comparisons between computations of the Navier-Stokes equations and linear theory in two- and three-dimensions can be found in Aalilija et al. \cite{aalilija2020analytical}. Following Lamb, they used the inviscid solution and an energy argument to find linear damping rates. For bouncing droplets,  Mol\'{a}\u{c}ek \& Bush considered a simplified model where both bath and droplet deform equally as a logarithmic spring \cite{molavcek2013drops}. Terwagne et al. experimentally investigated the role of deformation in the droplet rebound dynamics when the drop and bath are in different regimes, and proposed a double mass-spring system to replicate this range of regimes in a drop-bath system \cite{terwagne2013role}. While solid spheres can rebound in this regime \cite{galeano2021capillary}, droplet deformation converts kinetic energy to drop deformations on contact with the bath, and then may revert it back to kinetic energy at take-off, increasing upwards propulsion.

Non-dimensional parameters play a key role in quantifying the relative importance of physical mechanisms in the flow. For rebounding droplets and the bath deformation, the three parameters of particular interest are the Weber, Reynolds, and Ohnesorge numbers. Which for a typical impact velocity $W_0$, length scale of the unperturbed drop radius $R_0$, dynamic viscosity $\mu$, surface tension $\sigma$, and density $\rho$, are given as

\begin{equation*}
   W\!e = \frac{\rho W_0^2 R_0}{\sigma}, \qquad Re = \frac{\rho W_0 R_0}{\mu} ,\qquad Oh = \frac{\mu}{\sqrt{\rho \sigma R_0}} =  \frac{\sqrt{W\!e}}{Re}, 
\end{equation*}

where here the Weber number indicates ratio of inertia to surface tension, the Reynolds number encapsulates the relative effect of inertia to viscosity, and the Ohnesorge number provides a relationship between viscosity and surface tension. There are several other parameters which could play a role; in particular for larger drops and longer waves the Bond number $Bo = \rho g R_0^2/\sigma$ gains importance. Even small changes to these parameters will affect the overall droplet behaviour, including whether the drops will bounce, or coalesce \cite{lewin2024collision}. The work presented within this paper will focus around the low-Weber and the quasi-inviscid low-Ohnesorge limit, in order to capture smooth rebound dynamics with small disturbances \cite{sprittles2024gas}.

The investigation of single rebounds in this regime has thrived over recent years, with the main goal being computationally efficient models that faithfully reproduce the experimental and DNS results. 
The `kinematic match' method for axisymmetric \cite{galeano2017non} and non-axisymmetric impacts \cite{galeano2019quasi} modelled the droplet as a sphere, with the pressure on its contact surface obtained as a constraint for matching the shape of the sphere's and the surface of the bath. This method captured details of bouncing and walking droplets, matching experiments, and was later used in comparisons of solid sphere impacts \cite{galeano2021capillary}. The model neglected droplet deformation and the air layer. Alventosa \textit{et al.} \cite{alventosa2023inertio} use a simplified kinematic match methodology (matching the impact position at a single point) coupled with a deformable drop to model axisymmetric rebounds. They obtained good agreement with experiments and DNS, enabling a unified understanding of the bouncing dynamics in their target inertio-capillary regime (vanishing influence of viscosity and gravitational forces described by sufficiently small Ohnesorge and Bond numbers, i.e. $\text{Oh} \ll 1$ and $\text{Bo}\ll 1$).

These methods do not attempt to capture the behaviour of the lubricating air layer during the rebound dynamics. In this work we seek a coupled system allowing for the inclusion of the lubricating air layer in the rebound dynamics of a two-dimensional drop off a deep liquid bath in the low-Weber, low-Ohnesorge regime. We derive a closed set of evolution equations that govern the evolutionary dynamics of the droplet, free surface together with a free-boundary elliptic problem for the air layer pressure. This methodology aims to fill the gap between the kinematic match methods and  DNS of the full fluid equations. The system presented herein can be simulated much more efficiently than full DNS, whilst still providing key insight into the influence of the air gap, as we will showcase in the sections to follow.

We next present the derivation of the system, considering each fluid region independently in Section \ref{sect:Model}. We then couple the regions together and present a numerical approach for simulating the results in Section \ref{sect:Numerics}. Finally, we will use DNS techniques that have been previously verified against experiments to determine the accuracy of our model, and demonstrate that it can produce efficient simulations of complex multiple-rebound dynamics off a vibrating bath in Section \ref{sect:Results} and Section \ref{sect:Faraday}, respectively. We conclude with a discussion of the model and consideration for future avenues of study in Section \ref{sect:Conclusion}.
\section{Mathematical model}\label{sect:Model}

Consider a two-dimensional liquid droplet, as though an infinite-cylinder in the out-of-plane direction, moving through an air-filled upper half plane towards a deep bath of infinite depth and width as shown in figure \ref{fig:figure1}. We restrict the centre of mass of the droplet to vertical motion along the $z$-axis, introducing a symmetry to the problem with respect to the $z$ axis. Introducing the subscript notation $\cdot_\alpha$ to denote the different fluid regions, where $\alpha \in \{a,b,d\}$ are the air, bath, and droplet respectively, the velocity of each fluid is given by $\mathbf{u}_\alpha$, the pressure is given by $p_\alpha$, and $\mu_\alpha$, $\rho_\alpha$ and  $g$ denote the dynamic viscosity, density and gravity, respectively.

\begin{figure}
    \centering
    \scalebox{0.7}{

\tikzset{every picture/.style={line width=0.7pt}} 

\begin{tikzpicture}[x=0.75pt,y=0.75pt,yscale=-0.9,xscale=0.9]

\draw   (206,128) .. controls (206,61.73) and (259.73,8) .. (326,8) .. controls (392.27,8) and (446,61.73) .. (446,128) .. controls (446,194.27) and (392.27,248) .. (326,248) .. controls (259.73,248) and (206,194.27) .. (206,128) -- cycle ;
\draw  [draw opacity=0] (436.94,265.41) .. controls (411.34,298.61) and (371.17,320) .. (326,320) .. controls (279.63,320) and (238.52,297.46) .. (213.05,262.73) -- (326,180) -- cycle ; \draw   (436.94,265.41) .. controls (411.34,298.61) and (371.17,320) .. (326,320) .. controls (279.63,320) and (238.52,297.46) .. (213.05,262.73) ;  
\draw   (436,266.88) .. controls (436.54,266.26) and (437.08,265.63) .. (437.62,265) .. controls (444.22,257.32) and (450.54,250) .. (457.87,250) .. controls (465.2,250) and (471.52,257.32) .. (478.12,265) .. controls (484.72,272.68) and (491.04,280) .. (498.37,280) .. controls (505.7,280) and (512.02,272.68) .. (518.62,265) .. controls (525.22,257.32) and (531.54,250) .. (538.87,250) .. controls (546.2,250) and (552.52,257.32) .. (559.12,265) .. controls (565.72,272.68) and (572.04,280) .. (579.37,280) .. controls (581.67,280) and (583.86,279.28) .. (586,278.06) ;
\draw    (326,128) -- (326,188) ;
\draw [shift={(326,128)}, rotate = 90] [color={rgb, 255:red, 0; green, 0; blue, 0 }  ][fill={rgb, 255:red, 0; green, 0; blue, 0 }  ][line width=0.75]      (0, 0) circle [x radius= 3.35, y radius= 3.35]   ;
\draw    (326,128) -- (434.18,177.17) ;
\draw [shift={(436,178)}, rotate = 204.44] [color={rgb, 255:red, 0; green, 0; blue, 0 }  ][line width=0.75]    (10.93,-3.29) .. controls (6.95,-1.4) and (3.31,-0.3) .. (0,0) .. controls (3.31,0.3) and (6.95,1.4) .. (10.93,3.29)   ;
\draw  [draw opacity=0] (344.29,136.1) .. controls (341.2,143.08) and (334.23,147.96) .. (326.11,148) -- (326,128) -- cycle ; \draw   (344.29,136.1) .. controls (341.2,143.08) and (334.23,147.96) .. (326.11,148) ;  
\draw  (326,268) -- (356,268)(326,238) -- (326,268) -- cycle (349,263) -- (356,268) -- (349,273) (321,245) -- (326,238) -- (331,245)  ;
\draw  [dash pattern={on 0.84pt off 2.51pt}]  (586,268) -- (66,268) ;
\draw   (66,276.57) .. controls (67.83,277.48) and (69.72,278) .. (71.67,278) .. controls (79,278) and (85.32,270.68) .. (91.92,263) .. controls (98.52,255.32) and (104.84,248) .. (112.17,248) .. controls (119.5,248) and (125.82,255.32) .. (132.42,263) .. controls (139.02,270.68) and (145.34,278) .. (152.67,278) .. controls (160,278) and (166.32,270.68) .. (172.92,263) .. controls (179.52,255.32) and (185.84,248) .. (193.17,248) .. controls (200.5,248) and (206.82,255.32) .. (213.42,263) .. controls (214.28,264.01) and (215.14,265) .. (216,265.98) ;

\draw (337,146.4) node [anchor=north west][inner sep=0.75pt]  [font=\Large]  {$\theta $};
\draw (387.14,132.6) node [anchor=north west][inner sep=0.75pt]  [font=\Large,rotate=-23.8]  {$r$};
\draw (357,270.4) node [anchor=north west][inner sep=0.75pt]  [font=\Large]  {$x$};
\draw (327,222.4) node [anchor=north west][inner sep=0.75pt]  [font=\Large]  {$z$};

\end{tikzpicture}

}
    \caption{Schematic of model domain denoting two frame of reference points. The origin for the full frame is the Cartesian $(x,z)$-coordinate set directly beneath the south pole of the drop and aligned to an unperturbed bath free surface. The droplet will be evolved in a moving frame of reference with polar coordinates ($r,\theta)$ about the centre of mass of a drop whose free surface lies at $r=R_0$, and $\theta=0$ pointing vertically downwards towards the bath. }
    \label{fig:figure1}
\end{figure}
While each region of fluid will be treated differently due to particular approximations and geometries, in each region, $\mathbf{\Omega}_\alpha$, we begin with the incompressible Navier-Stokes equations

   \begin{align}\label{eq: mass}
    \nabla \cdot \mathbf{u}_\alpha &=  0, \qquad &\mathbf{x} &\in \Omega_\alpha,\\ \label{eq: mom}
    \rho_\alpha \left(\partial_t \mathbf{u}_\alpha + (\mathbf{u}_\alpha\cdot \nabla)\mathbf{u}_\alpha \right) &= -\nabla p_\alpha + \mu_\alpha \nabla^2 \mathbf{u}_\alpha - \rho_\alpha g \mathbf{e}_z, \qquad & \mathbf{x} &\in \Omega_\alpha,
\end{align}
where $g$ is the gravitational constant.
At the interface between fluids, defined by $F_\beta(\mathbf {x},t)=0$, where $\beta$ takes on the values $b$ or $d$, for the bath-air and drop-air interfaces respectively, the kinematic and stress continuity boundary conditions and the velocity continuity respectively are given by 

\begin{align}\label{eq: KBC General}
    \partial_t F_\beta + (\mathbf{u}_\beta \cdot \nabla)F_\beta &= 0, \qquad &x\in \partial\Omega_\beta,\\
    p_\beta \mathbf{n}_\beta - \mu_\beta \mathbf{\tau}_\beta \mathbf{n}_\beta &= p_a \mathbf{n}_\beta - \mu_a \mathbf{\tau}_a  \mathbf{n}_\beta + \sigma_\beta \kappa_\beta \mathbf{n}_\beta, \qquad & x\in \partial\Omega_\beta,\label{eq:StressCont} \\
    \mathbf{u}_\beta &=  \mathbf{u}_a,  \qquad & x\in \partial\Omega_\beta,\label{eq:VelocityCont}
   \end{align}
   
where $\sigma$ is the surface tension coefficient, $\kappa$ denotes the curvature of the free surface whose unit normal vector is given as $\mathbf{n}_\beta$, and  $\tau = (\nabla \mathbf{u} + \nabla \mathbf{u}^T)$ is the rate of strain tensor. In the far field for the air and the bath, the velocities vanish, and pressures tend to their hydrostatic balance values. 
We shall now consider the flows within the bath, air, and liquid domains separately, using the relevant approximation the respective domain requires.
   
\subsection{Linear quasi-potential bath model}

Following methodology well established in the literature \cite{milewski2015faraday,galeano2017non,galeano2021capillary, alventosa2023inertio} to describe the waves generated by a bath-droplet interaction, we proceed by assuming a fluid bath with small free-surface perturbations from the equilibrium $z=0$  denoted by $z=\eta_b (x,t)$. As the liquid is incompressible we can take the velocity field of the bath as  $\mathbf{u}_b = (u_b, w_b) = \nabla \phi + \nabla^\perp \psi$, where $\nabla \phi$ denotes the irrotational velocity component, and $\psi$ a small vortical correction arising from viscous effects. Here we have used the notation $\nabla^\perp = (-\partial_z, \partial_x)$. Due to the assumption that disturbances are small, the system \eqref{eq: mass} -- \eqref{eq:VelocityCont} results in the following linearized problem

\begin{align}
\nabla^2 \phi &= 0, \qquad &z&\le 0, \label{eq: laplace}\\
\rho_b \partial_t \phi &= - p_b - \rho_b g z, \qquad &z&\le 0, \\
\rho_b \partial_t \psi &= \mu_b \nabla^2 \psi, \qquad &z& \le 0, 
\end{align}
and boundary conditions

\begin{align}
\partial_t \eta_b  &= \partial_z \phi + \partial_x \psi , \qquad &z&=0,  \\
 p_b - p_a - \sigma_b \kappa_b &= 2\mu_b\left(\partial_z^2 \phi + \partial_{xz} \psi \right), \qquad &z&=0, \\
   \mu_a (\partial_z u_a + \partial_x w_a) &= \mu_b (2\partial_{xz} \phi + \partial_x^2 \psi - \partial_z^2 \psi) , \qquad &z&=0, \\
   u_a &= \partial_{x} \phi - \partial_z \psi, \qquad &z&=0, \\
    w_a &= \partial_{z} \phi + \partial_x \psi, \qquad &z&=0, \label{eq: z velocity match} \\
\phi,\psi &\to 0, \qquad &z &\to -\infty, \label{eq: end of full system}
\end{align}
which include the kinematic boundary condition, normal and tangential stress balances, velocity continuity, and far field decay, respectively. We have omitted normal stresses in the air, in anticipation of a lubrication layer approximation.

The arguments in Dias \textit{et al} \cite{dias2008theory} (building on the viscous damping rates for water waves obtained by Lamb \cite{lamb1924hydrodynamics}) involve a smallness assumption in the boundary layer flow represented by $\psi$ and a boundary layer scaling for $\psi$ in its vertical dependence. This is explained in more detail in Appendix \ref{App: Boundary}. The boundary conditions then simplify to

\begin{align}
\partial_t \eta_b  &= \partial_z \phi + \partial_x \psi , \qquad &z&=0,  \\
\partial_t \phi &=  - \frac{1}{\rho_b} p_a + g \eta_b + \frac{\sigma_b}{\rho_b} \partial_x^2 \eta_{b} - 2\nu_b \partial_z^2 \phi, \qquad &z&=0, \\
    \partial_t \psi &=  \nu_b \left(2+\frac{\mu_a}{ \mu_b}\right) \partial_{zx} \phi + \nu_b \frac{\mu_a}{\mu_b}\partial_z u_a , \qquad &z&=0, \\
   u_a &= \partial_{x} \phi, \qquad &z&=0, \\
    w_a &= \partial_{z} \phi, \qquad &z&=0,
\end{align}
where we have introduced  the kinematic viscosity $\nu_b = \mu_b/\rho_b$ and substituted for the linearized curvature of the surface $\kappa = -\partial_x^2 \eta_b$. Disregarding terms of $O({\mu_a}/{\mu_b})$, one can deduce using the leading order balance $\partial_t \eta_b = \partial_z \phi$ that $\partial_x \psi = 2\nu_b \partial_{x}^2 \eta_b$, and hence the boundary conditions for the velocity potential and the free surface evolution are forced by air pressure alone

\begin{align}
\partial_t \phi &= -\frac{1}{\rho_b} p_a -  g\eta_b + \frac{\sigma_b}{\rho_b}\partial_x^2 \eta_b  + 2\nu_b \partial_x^2 \phi, \qquad &z&=0, \label{eq: qp DBC} \\
\partial_t \eta_b  &= \partial_z \phi + 2\nu_b \partial_x^2 \eta_b, \qquad &z&=0, \label{eq: qp KBC}
\end{align}
where $\phi$ satisfies Laplace's equation in the lower half plane. The conditions for velocity continuity at the interface will be matched in the equations for the lubrication air layer, and the pressure $p_a$ will be taken to be zero outside this lubrication layer. We note that the kinematic boundary condition requires $\partial_z \phi$ to be evaluated at $z=0$, and that this can be expressed simply in terms of $\phi$ at the surface once the system is Fourier transformed in $x$ as shown in Section \ref{sect:Numerics}.
\subsection{Lubricating air layer model}

When the droplet of undisturbed radius $R_0$ is sufficiently far from the bath, the two liquid regions have a negligible effect on each other. However, when the height between the drop and the bath is of $\mathcal{O}(\varepsilon R_0)$, where $\varepsilon \ll 1$, they can start to affect each other and we are able to proceed by considering the air layer and the pressure within as a lubrication region \cite{hicks2013liquid}. 

The geometry of this lubrication region is given by
\begin{equation*}
\Omega_{l^*}(t) = \left\{(x,z): x \in [-l^*(t),l^*(t)], z \in [\eta_b(x,t),\eta_d(x,t)] \right\},    
\end{equation*}
where $l^*$ is the horizontal location at which the vertical distance between the bath and droplet becomes greater than $\varepsilon R_0$ for some chosen $\varepsilon \ll 1$. In the next section when we consider the evolution of the droplet it will be more appropriate to set $\eta_d = Z(t)-S(x,t)$, where $Z(t)$ is the vertical height of the centre of mass of the drop, and $S(x,t)$ is the shape profile of the lower interface of the droplet relative to that centre of mass. For rigid objects $S$ is assumed to be constant in time. The height of the lubrication layer within its horizontal domain, is given by 
\begin{equation}
    h(x,t) = \eta_d - \eta_b.
\end{equation}

Taking this geometry into consideration, and assuming weak impacts in which the lubrication layer will not deviate substantially from the horizontal, \eqref{eq: mass}-\eqref{eq: mom} can be reduced  with lubrication scaling to

\begin{align}
    \partial_x u_a + \partial_z w_a &= 0\label{eq: air mass}, \\
     0 &= -\partial_x p_a + \mu_a \partial_z^2 u_a,\label{eq: uzz} \\
    0 &= -\partial_z p_a. \label{eq: pz}
\end{align}
From equation \eqref{eq: pz}, we see that the pressure within the air layer varies in the horizontal direction only, i.e $p_a = P(x,t)$. At the two air-liquid interfaces we impose velocity matching as in \eqref{eq:VelocityCont}, such that

\begin{equation*}
    z=\eta_b: \quad u_a = \partial_x \phi, \quad w_a  =\partial_z \phi;  \qquad  \qquad z=\eta_d: \quad
    u_a = u_d, \quad w_a = w_d,
\end{equation*}
where we recall that $(u_d,w_d)$ are the velocity components of the flow in the droplet. Equation \eqref{eq: uzz} can then be solved to determine the form of $u_a$ as a Poiseuille flow within the air layer of the form 

\begin{equation}\label{eq: u_a}
    u_a = \frac{\partial_x P}{2\mu_a}(z-\eta_b)(z-\eta_d) + \frac{u_d|_{\eta_d}}{\eta_d- \eta_b}(z-\eta_b) + \frac{\partial_x \phi_b|_{\eta_b}}{\eta_b- \eta_d}(z-\eta_d).
\end{equation}

The lubrication equation results from integrating the conservation of mass equation \eqref{eq: air mass} across the air layer, 

\begin{equation}
    \int_{\eta_b}^{\eta_d} \partial_x u_a + \partial_z w_a \; \text{d}z = 0.
\end{equation}
Using the Leibniz integral rule 

\begin{equation}
    \partial_x \int_{\eta_b}^{\eta_d} u_a \text{d}z + \left(w_a\Big|_{\eta_d} - u_a \Big|_{\eta_d}\partial_x \eta_d\right)  - \left(w_a\Big|_{\eta_b} - u_a \Big|_{\eta_b} \partial_x \eta_b \right) = 0,
\end{equation}
which is a conservation law for the height of the layer, since the first term represents the area flux, and the boundary terms yield, from \eqref{eq: KBC General}, the change in thickness i.e. $\partial_x Q + \partial_t \eta_d  - \partial_t \eta_b = 0$ with a flux from (\ref{eq: u_a})

\begin{equation}
Q = -\frac{\partial_x P}{12\mu_a} h^3 + \frac{(\partial_x \phi_b|_{\eta_b} + u_d|_{\eta_d})}{2} h~.
\end{equation}
Hence the lubrication equation is

\begin{equation}\label{eq: thin film}
\partial_t h +  \partial_x \left[-\frac{1}{12\mu_a} \partial_x P h^3 + \frac{(\partial_x \phi_b|_{\eta_b} + u_d|_{\eta_d} )}{2} h \right] = 0.
\end{equation}
The boundary conditions for this equation are 
$P=0$ at $x=\pm l^*$.
\subsection{Droplet behaviour}

\subsubsection{Droplet trajectory}
The vertical motion of the droplet can be described using its centre of mass $Z(t)$. We shall ignore the air effects (primarily viscous drag) except for the interaction with the lubrication layer. Taking the mass of the droplet to be given by $\rho_d \pi R_0^2$, where $R_0$ is the undeformed radius of the drop, we write the vertical acceleration of the droplet as 

\begin{equation}\label{eq: vert dyn}
    \rho_d \pi R^2 \frac{\text{d}^2 Z}{\text{d}t^2} = \int_{-l^*}^{l^*} P dx - \rho_d \pi R^2 g,
\end{equation}
where the integral is the force due to the pressure from the thin film region, and $g$ is the gravitational constant.

If the droplet were to be modelled by a non-deformable disk we can take $h = \eta_d - \eta_b = (Z(t)-S(x))-\eta_b(x,t)$, with $u_d|_{\eta_d}=0$, and the system of equations consisting of (\ref{eq: laplace}), (\ref{eq: qp KBC}), (\ref{eq: qp DBC}), (\ref{eq: thin film}), and (\ref{eq: vert dyn}) is now closed.

\subsubsection{Droplet deformation}
The equations which govern the evolution of drop deformations follow the methodology first used by Lamb \cite{lamb1924hydrodynamics}, and more recently described by Aalilija \textit{et al} \cite{aalilija2020analytical} and Alventosa \textit{et al} \cite{alventosa2023inertio}. 
We describe the motion of fluid within the droplet in polar coordinates, with origin at the centre of mass of the drop and $\theta =0$ in the negative $z$-direction. The free surface of the droplet at rest lies along $r=R_0$, with small perturbations to this given by $\eta_d(\theta,t)$. 

In order to derive the equations one can re-cast the equations (\ref{eq: laplace})-(\ref{eq: z velocity match}) in polar coordinates, and proceed with a reduction for the viscous boundary layer near $r=R_0$ to obtain quasi-potential boundary conditions corresponding to (\ref{eq: qp KBC}) - (\ref{eq: qp DBC}). The resulting problem to solve is 

\begin{align}
 \nabla^2 \phi_d &= 0, \quad &r&<R_0 \\
    \partial_t \phi_d &= -\frac{1}{\rho_d}p_a - \frac{\sigma}{\rho_d} \kappa  - 2\nu_d\partial_r^2 \phi_d, \quad &r&=R_0 \label{eq: drop DBC}\\
    \partial_t \eta_d &= \partial_r \phi_d + 2 {\nu_d}\left(\frac{1}{r^2} \partial_\theta^2\eta_d - \frac{1}{r^3}\partial_\theta^2 \int \phi_d \;\text{d} t\right), \quad &r&=R_0, \label{eq: drop KBC}
\end{align}
where $\kappa = -\frac{1}{R^2_0}\left(\eta_d + \partial_\theta^2 \eta_d\right)$ is the linearized perturbation of curvature in polar coordinates. Solving Laplace's equation and expanding free surface deformations and pressure in terms of Fourier series symmetric about $\theta=0$, we require that

\begin{align}
    \phi_d(r, \theta, t) &= \sum_{n=2}^\infty a_n(t)r^n cos(n\theta), \\
    \eta_d(\theta, t) &= \sum_{n=2}^\infty c_n(t) cos(n\theta), \\
    p_a &= \sum_{n=2}^\infty \hat{p}_n(t) cos(n\theta),
\end{align}

where the $n=0$ terms are omitted to enforce mass conservation, and the $n=1$ terms are omitted as they correspond to the linearized center of mass evolution. Substitution into the expressions (\ref{eq: drop DBC})-(\ref{eq: drop KBC}) results in

\begin{align}
    \dot{a}_n R_0^n &= -\frac{1}{\rho_d}\hat{p}_n + \frac{\sigma}{\rho R_0^{2}} \left( 1-n^2\right)c_n - 2\nu_d \frac{1}{R_0^{2}}n(n-1)a_n R_0^n,\\
    \dot{c}_n &= \frac{1}{R_0} n a_n R_0^{n} + 2\nu_d \left(-\frac{1}{R_0^2}n^2 c_n  + \frac{1}{R_0^{3}} n^2  \int a_n R_0^n  \;\text{d}t\right). 
\end{align}
 
Considering the leading order balance $\dot{c}_n = \frac{1}{R_0} n a_n R_0^{n}$ into the viscous terms, and eliminating coefficients $a_n$ by differentiating the second equation and using the first equation, results in the evolution equation for the Fourier coefficient of the free surface shape 

\begin{equation}
    \ddot{c}_n + 2\lambda_n \dot{c}_n + \omega_n^2 c_n = -\frac{n}{\rho_d R_0}\hat{p}_n,
\end{equation}
where 

\begin{equation}
    \lambda_n = \frac{2 \mu_d}{\rho_d R_0^2}n(n-1), \qquad \omega_n^2 = \frac{\sigma_d }{\rho_d R_0^3}n(n^2-1) 
\end{equation}
are viscous damping and frequency terms, respectively. We note that these correspond to the values in Aalilija \textit{et al}\cite{aalilija2020analytical} using the energy method of Lamb\cite{lamb1924hydrodynamics}, but were obtained directly from the quasi-potential approximation. The surface tangential velocity which appears in the lubrication equation (\ref{eq: thin film}) can be recovered by the leading order 

\begin{equation}
    u_d(R_0, \theta, t) = -\sum_{n=2}^\infty \dot{c}_n(t) sin(n\theta).
\end{equation}

We have now closed the system of equations which describe fully the lubrication-mediated behaviour of a droplet rebounding off a deep liquid bath. Each of the three regions of fluid have been prescribed a governing set of equations, which are all coupled together through the presence of the pressure function $P(x,t)$. We will now present a numerical implementation of the system in Section \ref{sect:Numerics}.

\section{Numerical methods}\label{sect:Numerics}

 In the previous section we introduced the equations that govern each of the three fluid regions and how they fit within a full lubrication-mediated (LM) quasi-potential fluid model. We shall now introduce the numerical methods used to evolve this system during this section. We will also describe a high-accuracy direct numerical simulation (DNS) implementation for the full multi-fluid system, which can be used as a benchmark for the various approximations in our proposed methodology. Fully open-source codebases for the LM and DNS approaches presented in this section, as well as for the LM extension to Faraday systems detailed in Section~\ref{sect:Faraday}, can be found in the dedicated GitHub repository: \href{https://github.com/rcsc-group/BouncingDropletsLiquid2D}{https://github.com/rcsc-group/BouncingDropletsLiquid2D} \cite{openaccesscode}.

\subsection{Solving the lubrication-mediated quasi-potential model} 

The wave dynamics of the bath requires the solution of Laplace's equation in the lower half plane in order to compute the vertical fluid velocity $\partial_z \phi(x,z,t)|_{z=0}$, given $\phi(x,0,t)$, for the advection of the free surface. This is the simplest example of the use of Dirichlet-to-Neumann maps, and can be expressed succinctly in the Fourier domain. Defining $\Phi(x,t) = \phi(x,0, t)$, and denoting its Fourier transform as $\hat{\Phi}(k,t) = \mathcal{F}\{\Phi(x,t)\},$ the solution to Laplace's equation is given by 

\begin{equation*}
    \phi(x,z,t) = \mathcal{F}^{-1}\{e^{|k|z} \hat{\Phi}\},
\end{equation*}
where $\mathcal{F}^{-1}$ is the inverse Fourier transform. Hence, $\partial_z \phi(x,0,t) = \mathcal{F}^{-1}\{|k| \hat{\Phi}\}.$ The full lubrication-mediated quasi-potential model then can be written as a system of equations for $\Phi, \eta_b, \eta_d, Z, h, P$ as follows

\begin{align}
    \partial_t \Phi &= -\frac{1}{\rho_b}P - g\eta_b + \frac{\sigma_b}{\rho_b} \partial_x^2 \eta_b + 2\nu_b \partial_x^2 \Phi \label{eq: Numeric Phi_t},\\
    \partial_t \eta_b &= \mathcal{F}^{-1}\{|k|\mathcal{F}\{\Phi\}\} + 2\nu_b \partial_x^2 \eta_b, \\
    \ddot{c}_n &+ 2\lambda_n \dot{c}_n + \omega_n^2 c_n = -\frac{n}{\rho R_0} \hat{p}_n,
    \end{align}
   where
   
   \begin{equation}
   \eta_d=  R_0+\sum_{n=2}^\infty c_n cos(n\theta), \qquad \hat{p}_n = \frac{1}{\pi}\int_{-\theta^*}^{\theta^*} P cos(n\theta) \; d\theta, 
   \end{equation}
   and
   
\begin{equation}
    \frac{\text{d}Z}{\text{d}t} = W,  \qquad  \frac{\text{d}W}{\text{d}t} = \frac{1}{\rho_d \pi R_0^2}\int_{-l^*}^{l^*} P dx - g, 
\end{equation}
with 

\begin{equation}\label{eq: Numeric LubricationEquation}
    \partial_x \left[\frac{1}{12\mu_a} h^3 \partial_x P  - \frac{1}{2} h \partial_x \Phi \right] = \partial_t h, \qquad h = \left(Z-\eta_d\right) - \eta_b, \qquad x \in [-l^*,l^*].
\end{equation}
Mathematically, the above system should be considered as time-evolution differential equations for $\Phi, \eta_b, \eta_d, Z,$ together with a nonlinear elliptic problem (the lubrication equation) for $P$.

The lubrication region is bounded symmetrically by $|x|=l^*$ on the free surface and a corresponding $|\theta|=\theta^*$ on the droplet. Where necessary polar droplet coordinates and Cartesian free surface coordinates are related through standard transformations. The pressure outside of this lubrication region is considered to be zero. In order to define the edge of the lubrication region where $h \ll R_0$, we introduce $\varepsilon \ll 1$ such that $l^*$ solves

\begin{equation}
h(l^*) = \varepsilon R_0.
\end{equation}
In the numerical results we find that the lubrication layer thickness is insensitive to variation for a range of small values of $\varepsilon$, thus the value $\varepsilon=0.01$ is used to set the outermost point in the lubrication layer according to the formula above. In Moore \cite{moore2021introducing} the nondimensional value $(\mu_a/\rho_b R_0 W_0)^{1/3}$ obtained from  the horizontal momentum balance in the layer sets the vertical to horizontal scaling of the air layer. In our cases this value ranges between $0.05 < (\mu_a/\rho_b R_0 W_0)^{1/3} < 0.1$.

The system \eqref{eq: Numeric Phi_t}-\eqref{eq: Numeric LubricationEquation} is  spatially discretized on a periodic uniform grid spanning $[-L/2,\; L/2]$ using $N_x$ points, and using $N_\theta$ points around the droplet. We will choose a periodic domain in $x$ to approximate the problem, as this simplifies the evolution of the bath equations and the domain can be made sufficiently large such that waves from the boundary do not reach the droplet during impact. Typical numbers of grid points for the discretization were $N_\theta=2^8$ and $N_x = 2^{13}$ on a domain size of $L = 16 \pi R_0$. Further grid refinement did not change the numerical results appreciably, which was systemically tested through fixing a known stable set of parameters, and varying $N_\theta$ and $N_x$ both independently and concurrently. We varied $N_\theta$ between $2^6$ and $2^9$ in powers of 2, and a similar approach was taken to the bath for $N_x$ between $2^9$ and $2^{15}$, until the solutions converged.
Fast Fourier transforms were used for the evaluation of derivatives and the Dirichlet-to-Neumann map. The lubrication equation \eqref{eq: Numeric LubricationEquation} was integrated with the trapezium rule. Cubic splines were used for the interpolation of values from the polar grid on the droplet to the Cartesian grid on the bath.

Numerical time-stepping used the Matlab \texttt{ode45} adaptive Runge-Kutta scheme. We also used the basic Runge-Kutta $4^{\textrm{th}}$ order scheme which utilised fixed timesteps for some computations to verify the results through varying the timestep chosen until good agreement between models was met (which occurred for $\delta t = 10^{-9}$s) 
The adaptive timestepper was efficient as a small time step is needed during certain phases of impact: the system is highly sensitive at first contact between the drop and bath, and when the lubrication layer begins to retract. However, longer time steps can be taken during the remainder of the rebound, and a much larger ones prior to, or after contact. We have found that the droplet deformation can create instabilities which result in unphysical contact, particularly at higher impact speeds and larger droplets. The instabilities can be avoided by adding artificial damping (through an increased viscosity) to the droplet equations during impact. Solid impacts did not suffer from this instability. 

The initial conditions of the system were taken as the bath and droplet free surfaces lying unperturbed with the centre of mass of the droplet at a height $2R_0$ above the undisturbed surface of the bath, with some initial (negative) velocity $W_0$. This pre-impact height was chosen to facilitate direct comparison with the direct numerical simulation framework, which does not have a fixed criterion for bath-droplet interactions, instead solving the multi-phase flow equations in both liquid and gas regions in full over the timescale of each numerical experiment. Our dedicated implementation and setup for this problem is described in the following subsection. 

\subsection{Direct numerical simulation}
The DNS framework for this study has been designed using the Basilisk open-source code infrastructure \cite{popinet2009accurate,popinet2015quadtree}, using a finite volume discretisation scheme and second order adaptive in time and space capabilities which benefit the multi-scale nature of the problem. The interplay between inertial, capillary and viscous forces make the target problem especially challenging, with Basilisk a particularly suitable choice in this context in view of its versatile setup \cite{popinet2018numerical}. Variations of our implementation have been used successfully in the past within an axisymmetric formulation as part of studies of impacting superhydrophobic solids \cite{galeano2021capillary} and, more recently, liquid drops \cite{alventosa2023inertio}, with the present implementation representing a two-dimensional counterpart constructed to enable reduced-dimensional model comparisons with the deformable droplet model proposed in Section~\ref{sect:Model}. While previous multi-methodological investigations contained rigorous validation efforts, we have established the robustness of our code in the present configuration as well. Both trajectory metrics and detailed information such as gas film thickness, anticipated to be of $\mathcal{O}(1)\ \mu$m based on the recent experimental work of Tang \textit{et al} \cite{tang2019bouncing}, have been monitored in view of reaching mesh independence to a suitable level of accuracy. This required the usage of a dedicated non-coalescence setup \cite{sanjay2023drop}, which ensures no artificial merging of the drop and pool occurs owing to grid resolution limitations. This required the testing of minimum cell sizes down to sub-micron scales.

The computational box is set up to measure $32R_0$ in each dimension, and is half-filled with liquid, leading to sufficiently large domains in terms of both depth and lateral extent to prevent bottom effects and outgoing wave reflections to affect the results of interest. Resolution level $12$, pertaining to a setting with $2^{12}$ cells per dimension for the minimum cell size, was found to be sufficient to ensure reliable results in our parameter regimes of interest following careful numerical experimentation. The adaptive refinement strategy focused gridnodes in regions with fluid-fluid interfaces, as well as regions with large changes in magnitude of the velocity components, ensuring sufficient resolution in regions around the drop, as well as near the impact region, and supporting the accurate capturing of the outgoing waves in the pool. We note to have also made use of an interpolation method employing a harmonic mean formulation for the viscosity in the design of our implementation, as well as interface filtering techniques, owing to the relatively large contrast of the physical properties between the fluids used which may generate undesired instabilities, particularly in a rapidly evolving multi-scale structure such as the one present here. This led to runs characterised by $\mathcal{O}(10^5)$ gridnodes (a highly significant, and indeed necessary, decrease from the $\mathcal{O}(10^7)$ uniform grid counterpart setup) and approximately $\mathcal{O}(10^2)$ CPU hours per run, with the adaptive timestepping becoming particularly relevant in the delicate stages in which the drop and the pool are in contact.

The above data-rich framework provides access to carefully resolved drop and bath interface coordinates, as well as other general information of interest such as velocities and pressure field data, which enables verification of key hypotheses and ensures the creation of a strong foundation for the deformable drop model comparisons, which are expanded upon in Section~\ref{sect:Results}.
\section{Results}\label{sect:Results}


We present the numerical results of the lubrication-mediated quasi-potential model for both solid and deformable droplets, and begin by introducing a single test case which sits well within a stable parameter regime, using this to discuss the nuances of the model and how it compares to the DNS results in the deformable scenario. We proceed through a parameter sweep, highlighting the range of the model and general trends of droplet trajectories observed within a two-dimensional rebound case, all of which we present in the absence of gravity, as it is difficult for the droplet to detach under the effects of gravity in the two-dimensional system due to the stronger relative influence in the infinite-cylinder regime than a true spherical case \cite{vella2007surface}. Finally, we discuss the extensions to more complicated rebounds that can also be obtained, such as sustained periodic bouncing behaviour in the presence of Faraday forcing.

\begin{figure}
    \begin{center}
        \includegraphics[width = \textwidth]{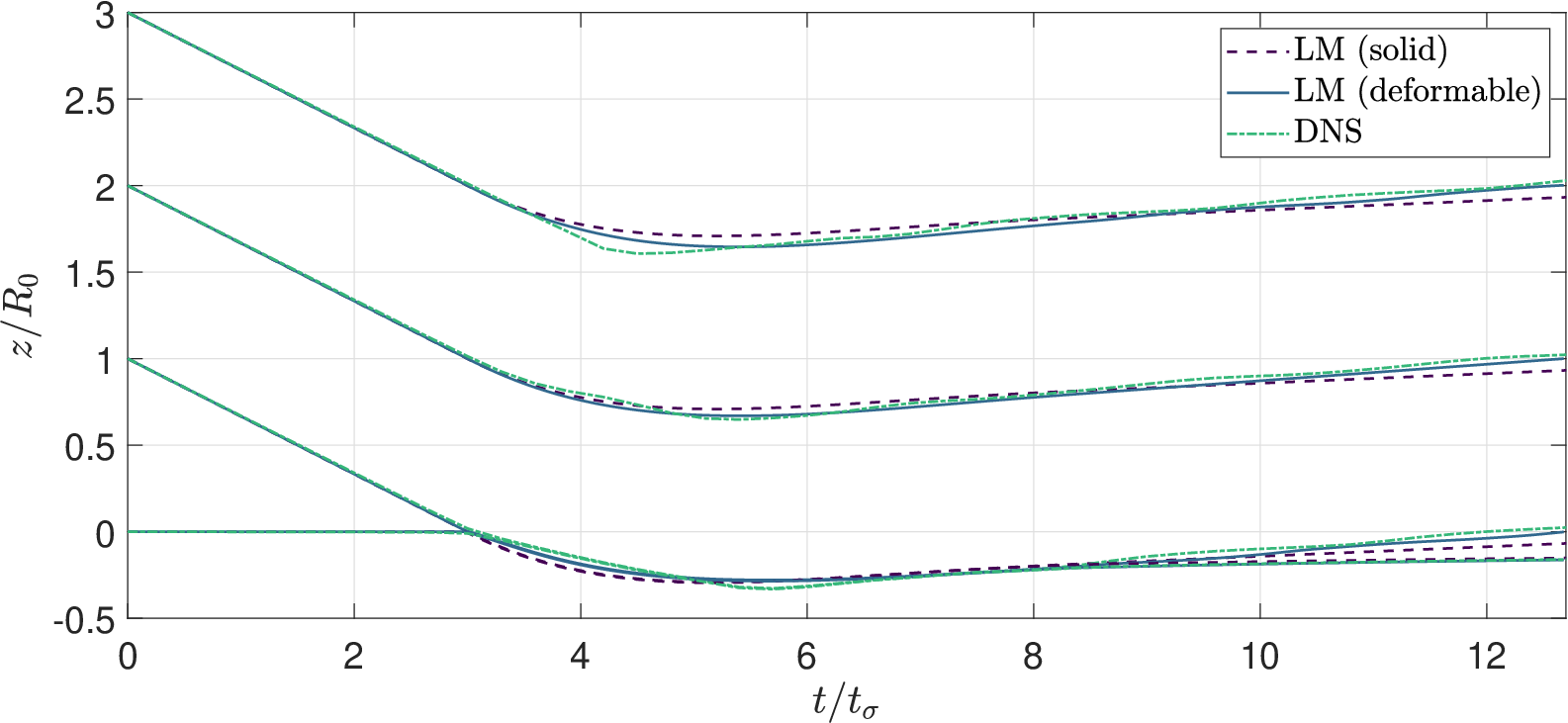}
        \caption{Comparison of lubrication-mediated model for the solid and liquid drop cases, and DNS results for a deformable droplet, with initial undeformed radius $R_0 = 0.2$mm released with velocity $W_0=-0.2$ms$^{-1}$ from a centre of mass  height of $2R_0$. Shown are the centre of mass and poles of both droplets, and the free surface directly beneath the south pole. Simulations are run until the centre of mass returns to a height $R_0$.The liquid drop has the physical properties of water, and the solid impactor has the same density, with all other parameters indicated in Table~\ref{tab: Appendix}.}\label{fig: Rebound Comparison (t,z)}
    \end{center}
\end{figure}

\subsection{LM model investigation}

Within Figures \ref{fig: Rebound Comparison (t,z)}-\ref{fig: h,Q,P nondimmed} we first consider a droplet with radius $R_0 = 0.2$ mm and initial velocity of $W_0 = -0.2$ms$^{-1}$, and compare the cases where the droplet is either solid or liquid to highlight the effect of droplet deformation on the rebound behaviour. In both cases we take the liquid in the bath to be water, and we take the droplet to be water or solid with the density of water. Parameter values are presented in Appendix~\ref{app: Table}. To consider solid droplets, the surface of the sphere is fixed and not permitted to deform setting $c_n=0$, hence $\eta_d(\theta,t) = R_0$. For further discussion of how viscous damping during impact affects rebound dynamics (including consideration of the soft solid limit), please see Appendix \ref{app: Visc}.

Figure~\ref{fig: Rebound Comparison (t,z)} presents the trajectories in the solid and deformable lubrication-mediated models, the latter also being contrasted to their counterpart DNS results. The lubrication-mediated rebound in the deformable case shows good agreement with DNS on the whole, though has a much smoother contact trajectory, which can be explained by the enhanced viscosity used during impact causing it to behave more closely to the solid drop during contact. Post contact, the droplet poles in both computations show oscillation generated by the deformation during contact, however the DNS oscillations are larger. The solid case has a slower exit velocity, due to the solid drop losing more kinetic energy to the bath, and not storing energy during impact as we shall see below. Henceforth, results are plotted nondimensionally using a capillary timescale $t_\sigma = \sqrt{\rho R_0^3/\sigma}$ and pressure $\sigma/R_0$ and with radius and initial velocity being used for typical length scale and velocity. 
We discuss the choice of enhanced viscous damping briefly in Appendix~\ref{app: Visc}.

Figure~\ref{fig: Energy Plots (t,E) } demonstrates the transfer and decay of energy in the system in both the solid and deformable rebound cases. We can write the energy balance as 
\begin{equation*}
    E_{\text{tot}}(t) = K\!E_{\text{drop}} + E_{\text{wave}}  + E_{\text{osc}} - \Delta E_{\text{damp}},
\end{equation*}
where $E_{\text{tot}}$ denotes the total energy of the system, which at the start of impact is stored entirely in the kinetic energy of the droplet $ K\!E_{\text{drop}}$. $E_{\text{wave}}$ is the energy within the bath's wave field and $E_{\text{osc}}$ is the energy within the oscillations in the droplet. 

\begin{figure}
    \begin{center}
        \begin{subfigure}{0.49\linewidth}
            \includegraphics[width = \textwidth]{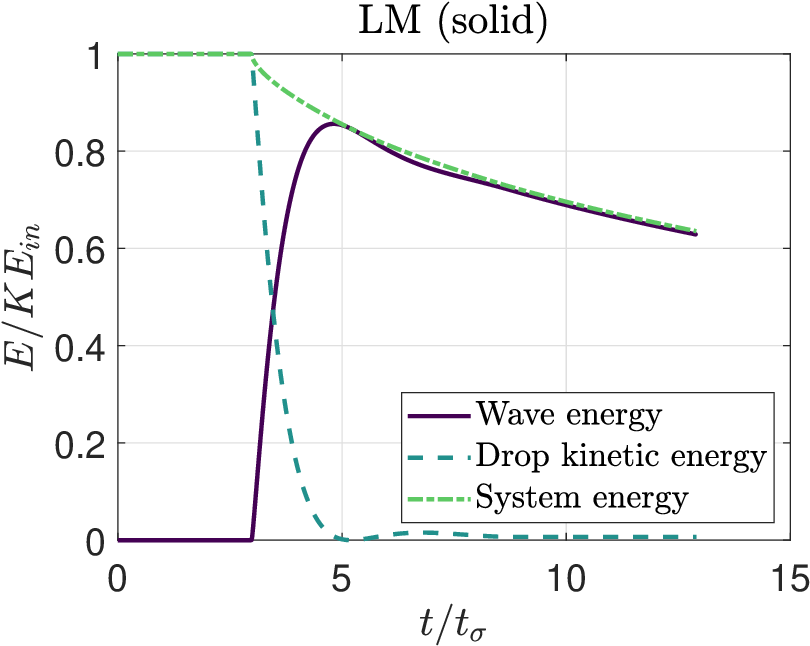}
        \end{subfigure}
        \hspace*{\fill}
        \begin{subfigure}{0.49\linewidth}
            \includegraphics[width = \textwidth]{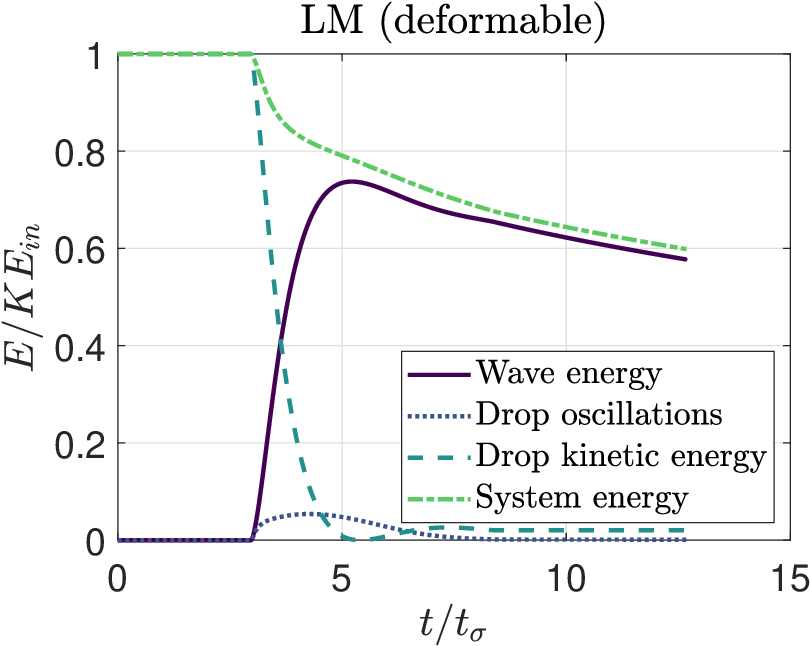}
        \end{subfigure}
        \caption{Energy balance for the solid and deformable impactors for the lubrication-mediated model, with parameters as shown in Figure \ref{fig: Rebound Comparison (t,z)}, and in particular initial radius $R_0 = 0.2$mm, and initial velocity $W_0=-0.2$ms$^{-1}$. The energy in the system starts in the droplet kinetic energy $K\!E_{\text{drop}}$, and is transferred to the deformation and waves on the bath, and, in the deformable case, to the droplet oscillations during impact. The system energy is computed to be the sum of these terms, from which we can the loss of system energy due to the dissipation within the system.} \label{fig: Energy Plots (t,E) }
    \end{center}
\end{figure}

Once impact begins energy is transferred to the free surface motion and (for deformable drops) to droplet deformation. The effect of drop deformation on the overall dynamics is clear: deformation lessens the energy transferred into the waves of the bath, instead transferring energy to the oscillations within the droplet. Some of the droplet energy is returned in a rebounding effect and contributes to the upwards velocity of the droplet, increasing its kinetic energy relative to the solid case. This results in the higher exit velocity seen in the deformable case despite a lower final total energy. The effect of viscous damping is seen through the loss of total energy which is enhanced in the deformable drop case. The Ohnesorge number appears naturally as a decay constant, with a typical decay over a capillary timescale of $\exp{(-2\pi Oh)} \approx 0.95$, matching qualitatively the computational results.

\begin{figure}
    \begin{center}
        \begin{subfigure}{0.48\linewidth}
            \includegraphics[width = \textwidth]{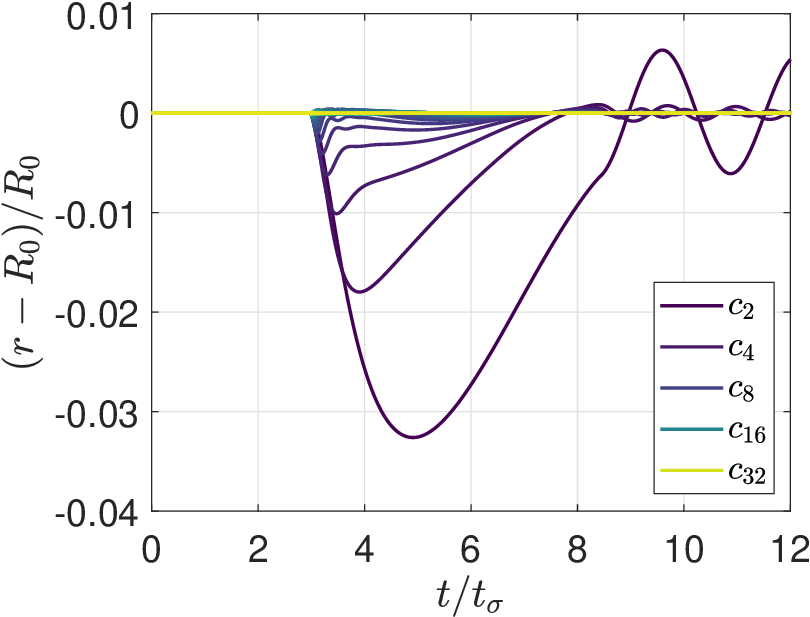}
            \subcaption{Fourier modes of the droplet free surface ($n=2,3,\ldots,31,32$). The $n=2$ mode experiences the greatest excitation and governs most of the droplet behaviour, with higher order modes experiencing less initial excitation and decaying more quickly. \\}
        \end{subfigure}
        \hspace*{\fill}
        \begin{subfigure}{0.48\linewidth}
            \includegraphics[width = \textwidth]{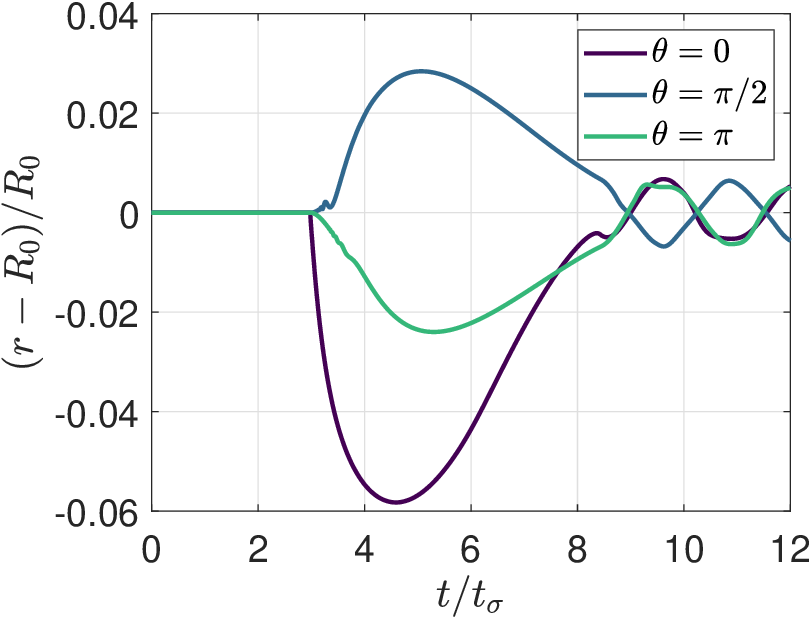}
            \subcaption{Movement of the poles and equator from the rest state of the droplet. The south pole at $\theta =0$  experiences the largest disturbance as the bottom of the droplet flattens, with widening evident at the equator $\theta =\pi/2$. Oscillations are clear after the initial impact. \\}
        \end{subfigure}
        \hfill
        \begin{subfigure}{\linewidth}
            \includegraphics[width = \textwidth]{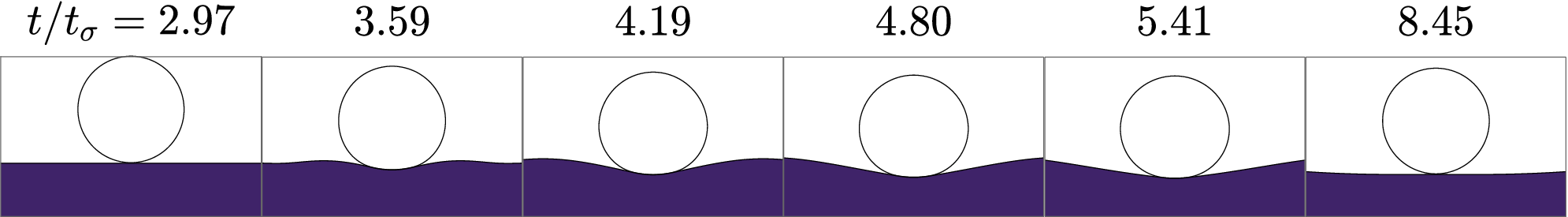}
            \subcaption{Snapshots of the deformable droplet rebound from point of contact at $t=t_{\text{imp}}$ to detachment at $t=t_{\text{liftoff}}$, including maximum deformation at 4.19 and maximum depth at 4.80. 
            }
        \end{subfigure}
        \caption{Droplet deformation and droplet rebound characteristics for the deformable droplet lubrication-mediated model, as shown in Figure \ref{fig: Rebound Comparison (t,z)}, with initial radius $R_0 = 0.2$mm, and initial velocity $W_0=-0.2$ms$^{-1}$.}\label{fig: Droplet Deformations}
    \end{center}
\end{figure} 


While the drop oscillation energy is small we may investigate the deformation with both the values of the Fourier coefficients of the drop's surface ($c_n$), and the overall shape profile of the droplet over time. Fig \ref{fig: Droplet Deformations}(a) illustrates normalised modes $c_n/R_0$, (where $n=2,\ldots, 32$) and it is clear that the second mode is the largest to be excited, and after detachment remains the main component of the oscillation. In contrast, for $n>6$ the modes have negligible effect on the overall behaviour post impact. They are however all excited at first contact, and we see this spike downwards as the southern half of the droplet essentially tries to `flatten' out as the lubrication layer initially spreads out. We see this in Figure~\ref{fig: Droplet Deformations}(b), where the south pole is deformed just before the north pole. Interestingly in this plot it is even possible to see the surface waves ripple around the droplet as they are excited at the equator and then at the north pole. Again, post impact we see the droplet continue to oscillate with period $t_\sigma$.

\begin{figure}
    \begin{center}
        \begin{subfigure}{0.49\linewidth}
            \includegraphics[width = \textwidth]{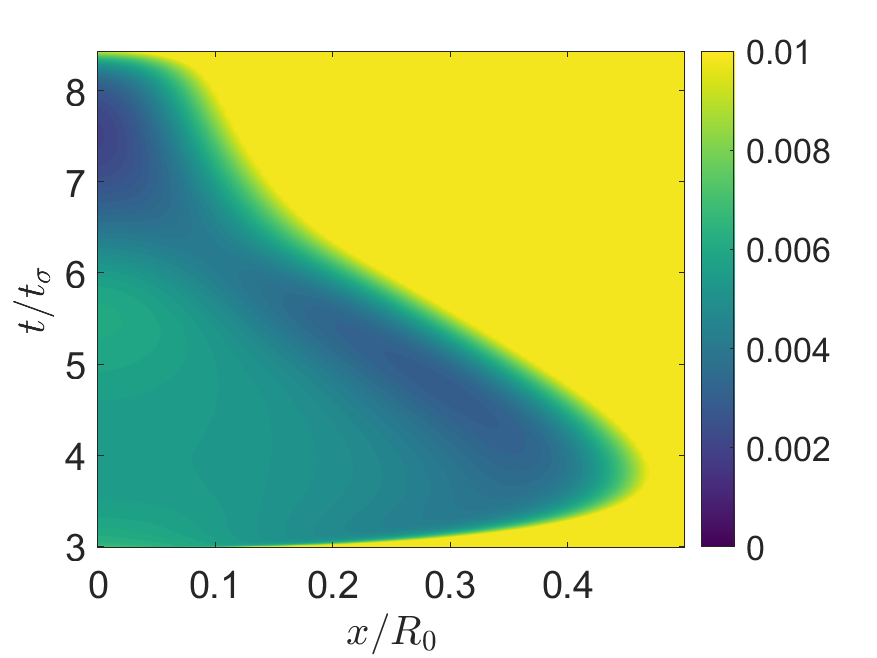}
        \end{subfigure}
        \begin{subfigure}{0.49\linewidth}
            \includegraphics[width = \textwidth]{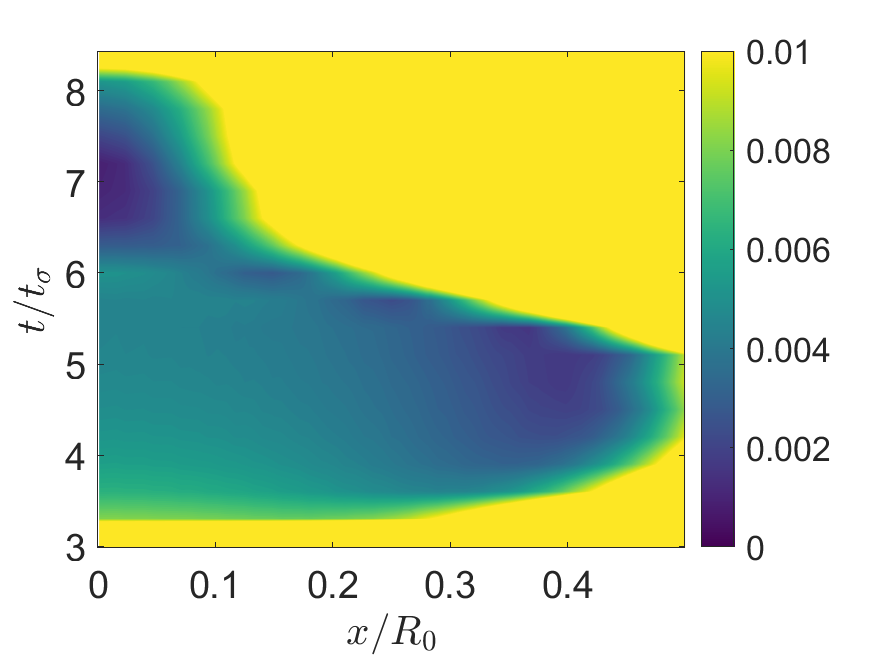}
        \end{subfigure}
        \caption{Dimensionless air layer height for a deformable droplet impact for the lubrication-mediated model (left) and DNS (right) as shown in Figure \ref{fig: Rebound Comparison (t,z)}, with initial radius $R_0 = 0.2$mm, and initial velocity $W_0=-0.2$ms$^{-1}$. The time axis in both figures is restricted to the approximate contact time, which begins from time $t/t_\sigma \approx 3$, to highlight impact behaviour. At the edges of the lubrication region and during the pinch before lift-off the dimensional film thicknesses may be as small as $ 0.5\;\mu$m.}\label{fig: Air Layer (x,t)}
    \end{center}
\end{figure}

A key distinction between the work presented here and the current literature is the dynamic inclusion of the lubrication layer  within the model. In Figure~\ref{fig: Air Layer (x,t)} we show a spatio-temporal plot of the lubrication layer height, over $\Omega_{l^*}(t)$. There are three main stages of rebound. The first stage happens quickly as the droplet ``impacts'' the bath and the lubrication region spreads out on timescales shorter than $t_\sigma$. Rapidly, the lubrication region  reaches its maximum extent and in the second stage begins to recede as the droplet peels away from the bath. During this stage, a narrow region forms near the outer edge where the layer depth is much thinner than at the centre of the lubrication region. If coalescence were permitted by the model, this would be the point 
of first contact that is described in the literature \cite{thoroddsen2003air}. We could expect to see contact when this depth is sufficiently small for atomic scale effects (e.g. Van der Waals forces) to become relevant. The third stage is pinchoff, which happens on a very short timescale and with negative pressure (suction) near the pinchoff point. 

\begin{figure}
    \begin{center}
        \begin{subfigure}{0.65\linewidth}
            \includegraphics[width = \textwidth]{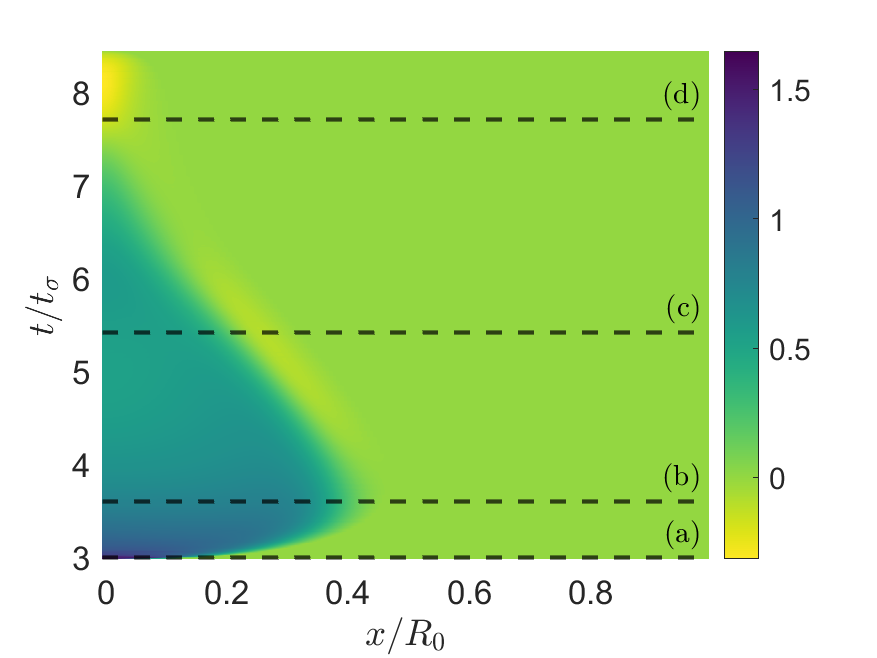} 
            \caption{}\label{fig: Pressure Surf}
        \end{subfigure}
        \begin{subfigure}{0.34\linewidth}
            \includegraphics[width = \textwidth]{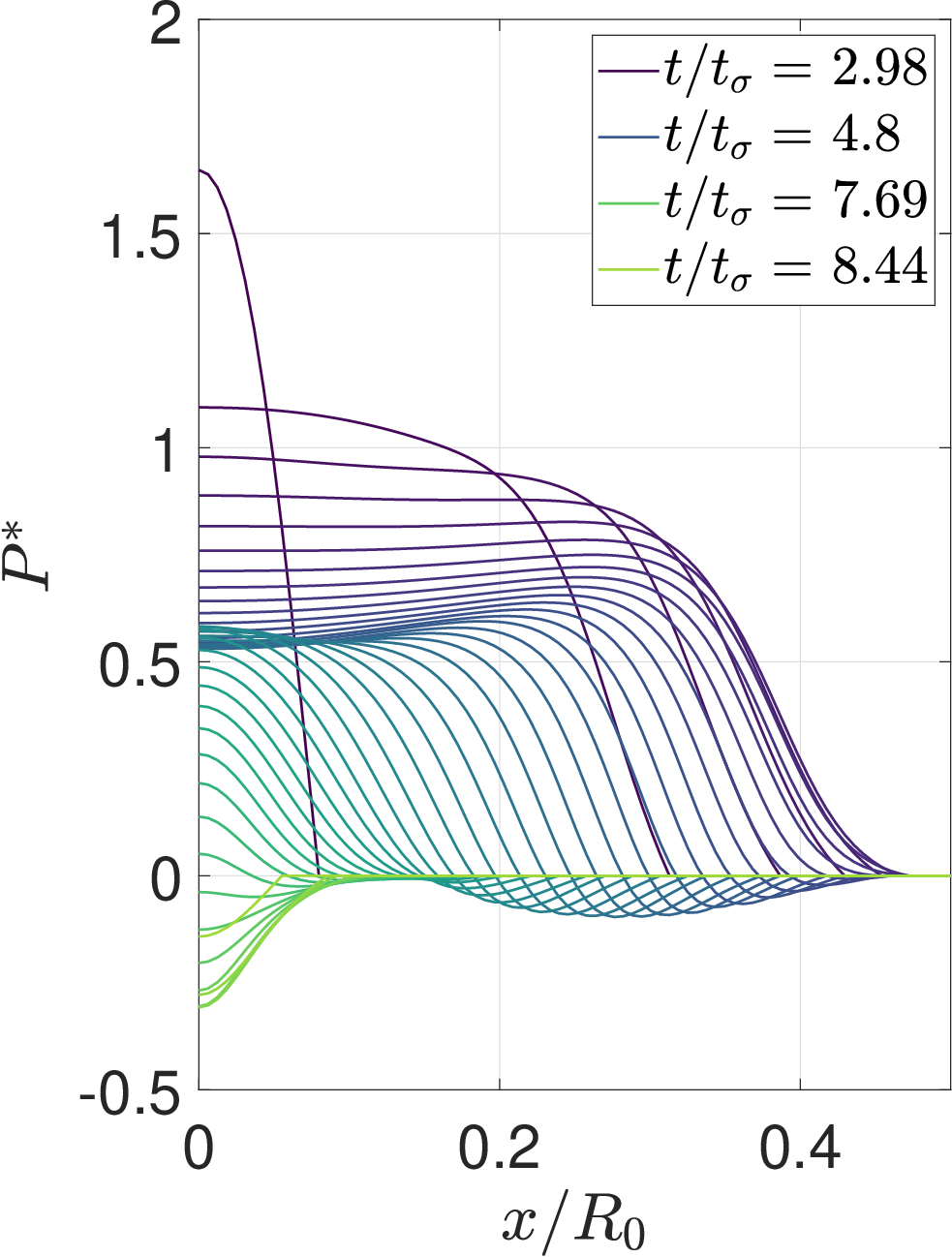}
            \caption{}\label{fig: Pressure Slices}
        \end{subfigure}
    \end{center}
    \caption{Dimensionless pressure $P^*$ within the air layer as determined by equation \eqref{eq: Numeric LubricationEquation} for the deformable lubrication-mediated rebound presented in Figure \ref{fig: Rebound Comparison (t,z)}, with initial radius $R_0 = 0.2$ mm and initial velocity $W_0 = -0.2$ms$^{-1}$, presented as (a) a spatio-temporal plot (with impact beginning from time $t/t_\sigma \approx 3$), and (b) at different times. The large spike of pressure at initial ``impact'' quickly spreads out and retains a flattened profile for most of contact. As the pressure diminishes in the later stages a region of low pressure forms on the boundary and propagates inwards, generating a suction effect. Eventually the suction region reaches the centre of the droplet, before detachment. The black lines in panel (a) correspond to the snapshots within figure~\ref{fig: h,Q,P nondimmed}. }\label{fig: Pressures}
\end{figure}

\begin{figure}
    \begin{center}
        \begin{subfigure}{0.49\linewidth}
            \includegraphics[width = \textwidth, left]{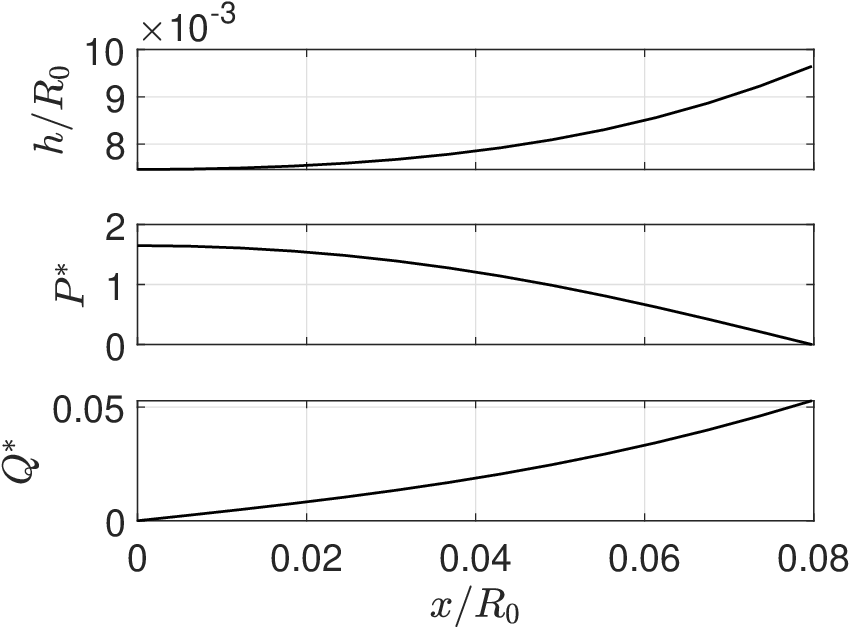}\caption{$t/t_\sigma=2.98$}
        \end{subfigure}
        \begin{subfigure}{0.49\linewidth}
            \includegraphics[width = \textwidth, left]{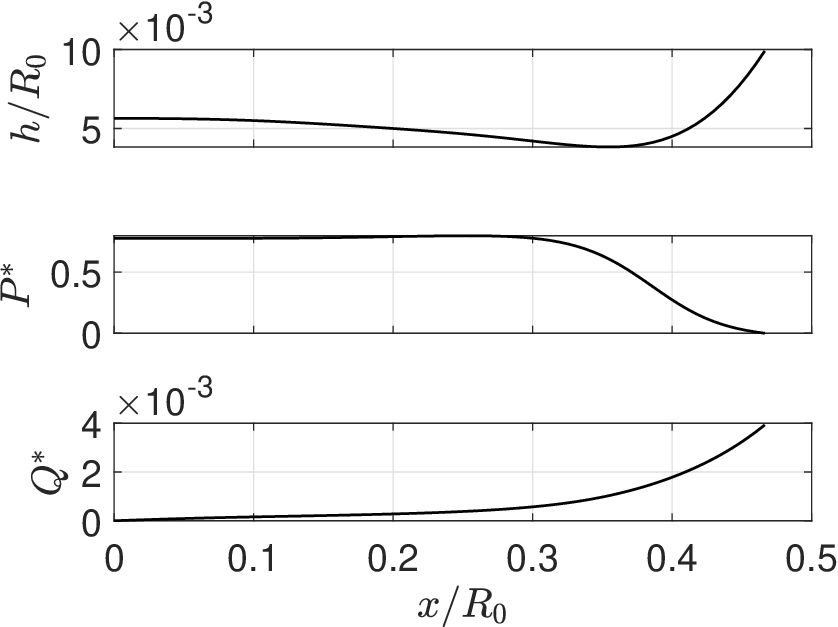}\caption{$t/t_\sigma=3.59$}
        \end{subfigure}
        \begin{subfigure}{0.49\linewidth}
            \includegraphics[width = \textwidth, right]{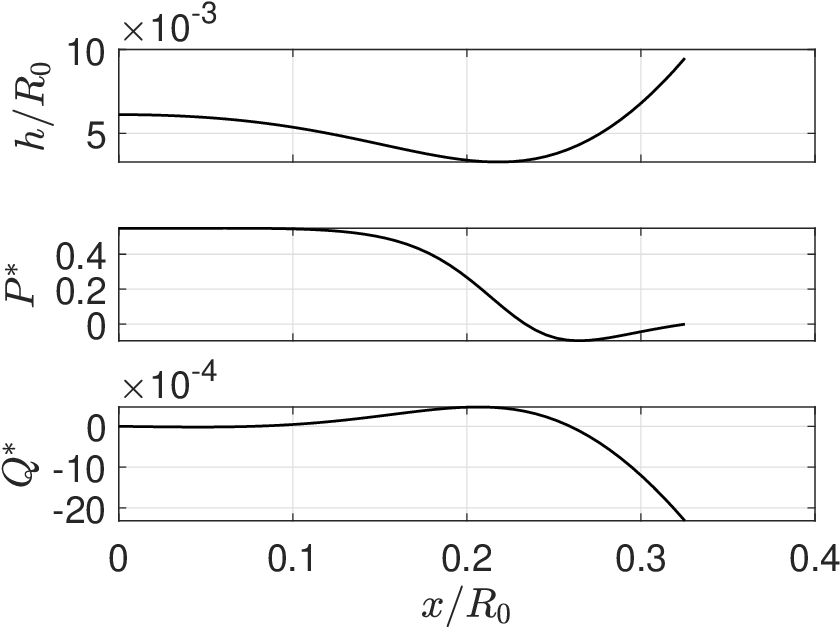}\caption{$t/t_\sigma=5.41$}
        \end{subfigure}
        \begin{subfigure}{0.49\linewidth}
            \includegraphics[width = \textwidth, right]{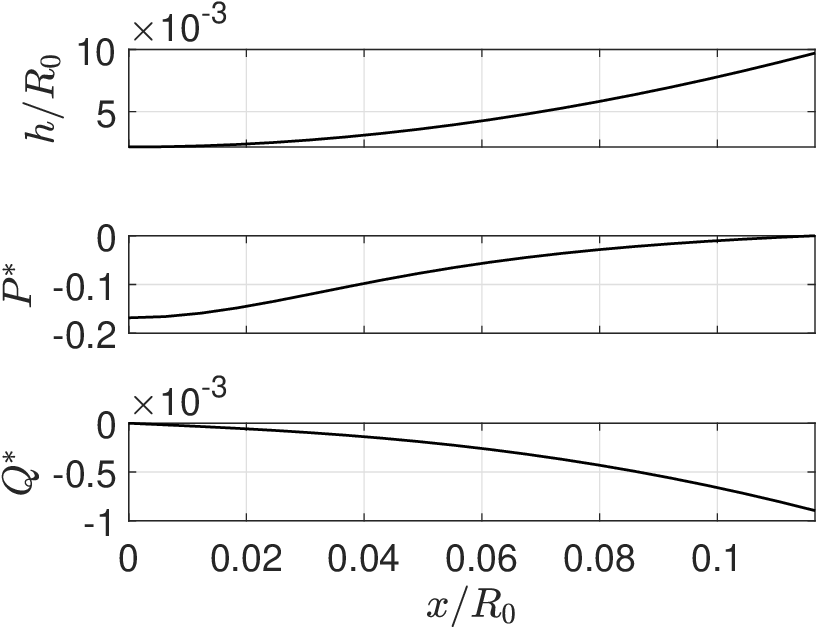}\caption{$t/t_\sigma=7.69$}
        \end{subfigure}
        \caption{Four representative snapshots of nondimensional layer height $h^*$, pressure $P^*$ and flux $Q^*$ within the lubrication region $\Omega_{l^*}(t)$ during the deformable lubrication-mediated rebound presented in Figure \ref{fig: Rebound Comparison (t,z)} with initial radius $R_0 = 0.2$mm and velocity $W_0 = -0.2$ms$^{-1}$. The times correspond to solid lines on Figure~\ref{fig: Pressure Surf}, chosen to highlight the change in behaviour between at distinct time regions during impact; initial contact (stage 1), widest point of contact, slow detachment (stage 2), and pinch-off (stage 3). Note that the axes change between each snapshot.}\label{fig: h,Q,P nondimmed}
    \end{center}
\end{figure}

The centre lubrication region in both the lubrication-mediated model and DNS presented is of near constant thickness of  $\mathcal{O}(1)\ \mu$m and pressure $\mathcal{O}(\sigma/R_0)$ for most of the impact, a quasi-static air cushion. This was experimentally observed in Tang \textit{et al} \cite{tang2019bouncing}, and corroborated in DNS computations \cite{galeano2021capillary, alventosa2023inertio, sprittles2024gas}. The region where the layer pinches at its boundaries lines up with the maxima (at the transition between stage 1 and stage 2) and minima (during stage 2) at the edge of the pressure profile. 

In Figure~\ref{fig: Pressures}(a) the horizontal bars across the plot correspond to the plots shown in Figure~\ref{fig: h,Q,P nondimmed}. There we show the nondimensional lubrication layer height $h/R_0$, pressure profile $P^*=P/(\sigma/R_0)$, and flux $Q^*$ against $x/R_0$, thus also indicating the fraction of the unperturbed radius they take up. These four times were selected as representative of impact stages. 

Figure~\ref{fig: h,Q,P nondimmed}(a) shows the profiles that the height, pressure, and flux during stage 1. The height of the layer is monotonically increasing, mimicking the shape of the droplet. There is a large spike in pressure, monotonically decreasing to atmospheric pressure, and a monotonically increasing flux as air is being forced out of the layer.

Figure~\ref{fig: h,Q,P nondimmed}(b) corresponds to the widest contact region, which occurs before the drop has reached maximum depth. The pressure is near constant (and hence flux is near zero) for the majority of the lubrication region with a very slight maximum near its edge and a smooth decay to atmospheric pressure at the outer region of the layer. The maximum is more pronounced in three-dimensions \cite{phillips2024lubrication}. The narrowing of the layer at its edge is now apparent.

Figure~\ref{fig: h,Q,P nondimmed}(c) corresponds to stage 2. The layer edge is propagating inwards towards the centre of the drop, the trapped air region contracts and the narrowing is more pronounced. There is a suction zone (negative pressure) and corresponding (negative) flux of outside air towards the outer boundary of the layer.

In the final stage 3 of impact (Figure~\ref{fig: h,Q,P nondimmed}(d)), the pinched regions on both sides of the drop have come together in the centre, resulting in a very thin layer there, and monotonically increasing away from the origin. The pressure is negative, delaying the detachment, and monotonically increasing. The flux corresponds to air filling the gap. Subsequently, the drop detaches and moves into flight away from the surface of the bath.

\subsection{Parameter study}

Through varying dimensionless grouping values $W\!e$ and $Oh$ in the absence of gravity, we construct an understanding of wider trends that can be observed. We present a series of cases within a region of applicability of the model, noting that the assumption that the deformations are small begin to break down at higher Weber numbers. 

\begin{figure}
    \begin{center}
        \includegraphics[width = \textwidth]{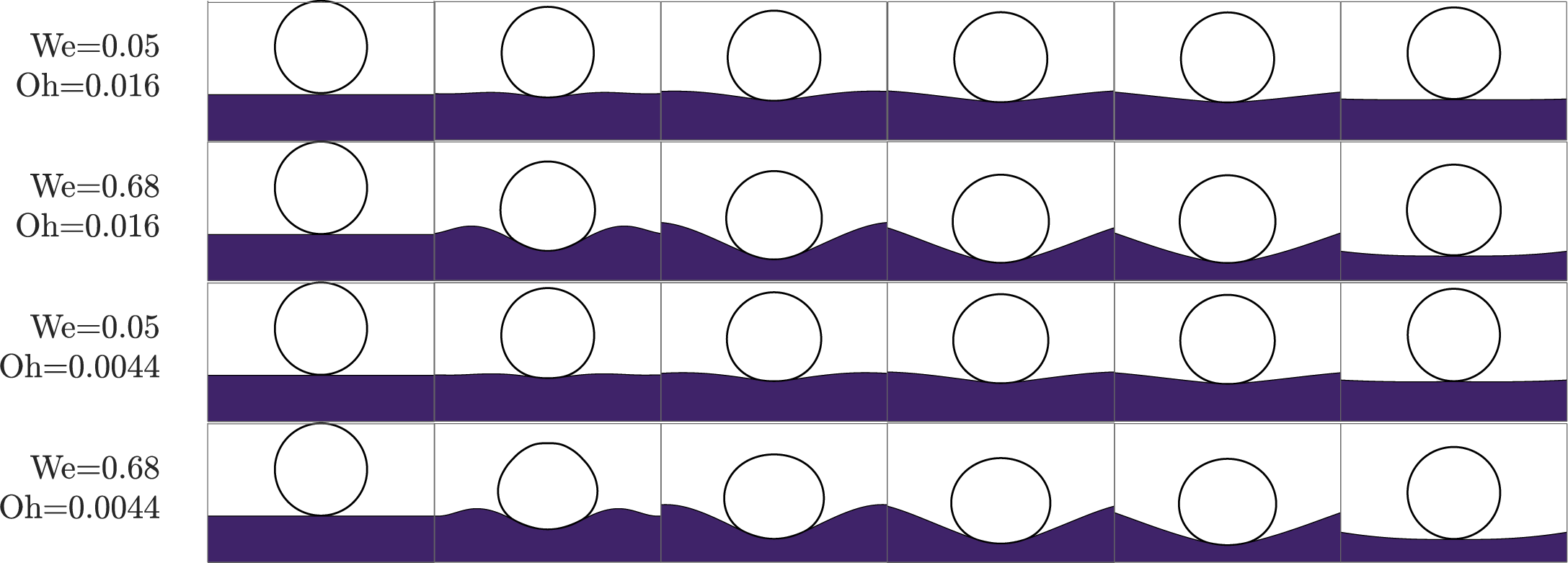}
        \caption{The evolution of four parameter regimes for the deformable LM model from point of impact to detachment. The top two runs correspond to a smaller droplet of radius $R_0=0.05$mm, and the lower two correspond to a larger droplet with radius $R_0=0.7$mm. The Weber number was varied by changing the initial velocity of the droplets respectively.}\label{fig: Deformation comparisons}
    \end{center}
\end{figure}

We begin this parameter study by first highlighting the extremal values the investigation, in particular looking at distinct $W\!e$ and $Oh$ runs in Figure~\ref{fig: Deformation comparisons}. Scaled to highlight differences in deformation, the figure shows the four cases at the same nondimensional times. The first snapshot is taken at $t_\text{imp}/t_\sigma$, and the last at $t_\text{detachment}/t_\sigma$, where we define $t_\text{detachment}$ as the time when the drop and bath detach as the distance between them is larger than the threshhold $\varepsilon R_0$. The middle columns showcase intermediate deformations, with the penultimate column corresponding to maximum depth. The figures highlight that deformation of both of the liquid interfaces is more strongly linked to the Weber number than the Ohnesorge number, however the balance between viscous and surface tension effects still plays a part, as the bottom row at largest $W\!e$ and smallest $Oh$ of the four rows shows greatest deformation in the drop and bath. The Reynolds number $Re=\sqrt{W\!e}/Oh$ increases from row 1 to row 4 and is approximately equal for the two middle rows. 

 In what follows, and to agree with prior studies, the time of impact $t_\text{imp}$ is defined as when the centre of mass of the droplet passes $z=R_0$ and $t_{\text{liftoff}}$ is when the centre of mass passes $z=R_0$ upwards again.  The contact time $t_c=t_{\text{liftoff}}-t_\text{imp}$ is the difference between these two times. This choice of time to determine contact will be different from when the lubrication layer `switches on' in the model, but this is done for a more accurate comparison to results in the wider literature. The other values we will consider are the maximum depth reached by the south pole of the droplet which we denote as $\delta$, and present as fraction of radius, and the coefficient of restitution, the negative of ratio of separation velocity at $t_\text{liftoff}$ to contact velocity at $t_\text{imp}$. Summarising:
\begin{equation*}
   \alpha = -\frac{W_{\text{out}}}{W_{\text{in}}}, \qquad \delta = \frac{(Z_{min}-R_0)}{R_0} ,\qquad \tau = \frac{t_c}{t_\sigma},
\end{equation*}
where $Z_{min}$ is the lowest point the centre of mass achieves during impact.
\begin{figure}
    \begin{center}
        \begin{subfigure}{0.49\linewidth}
            \caption{}
            \includegraphics[width = \textwidth]{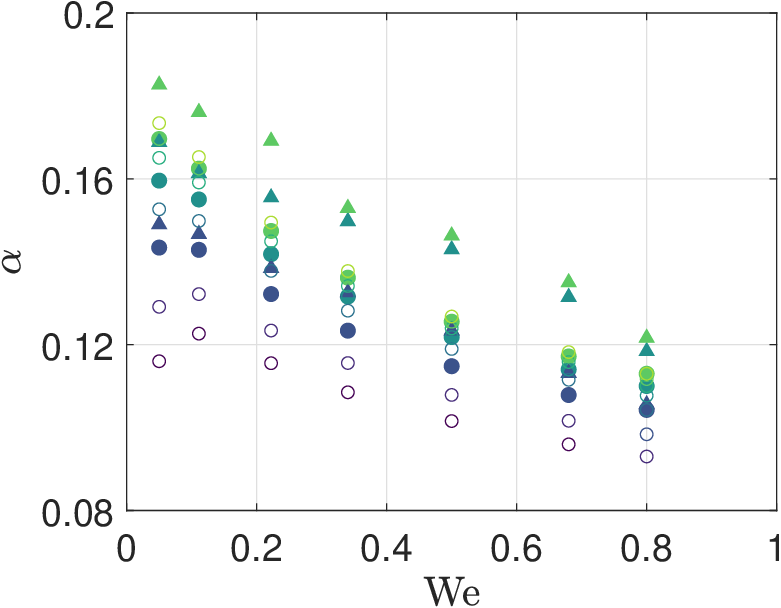}
        \end{subfigure}
        \begin{subfigure}{0.49\linewidth}
        \caption{}
            \includegraphics[width = \textwidth]{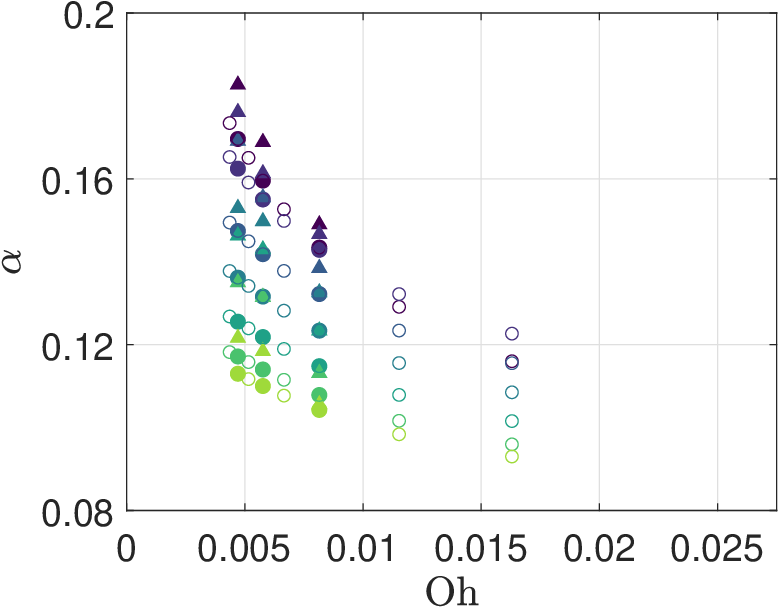}
        \end{subfigure}
        \begin{subfigure}{0.49\linewidth}
        \caption{}
            \includegraphics[width = \textwidth]{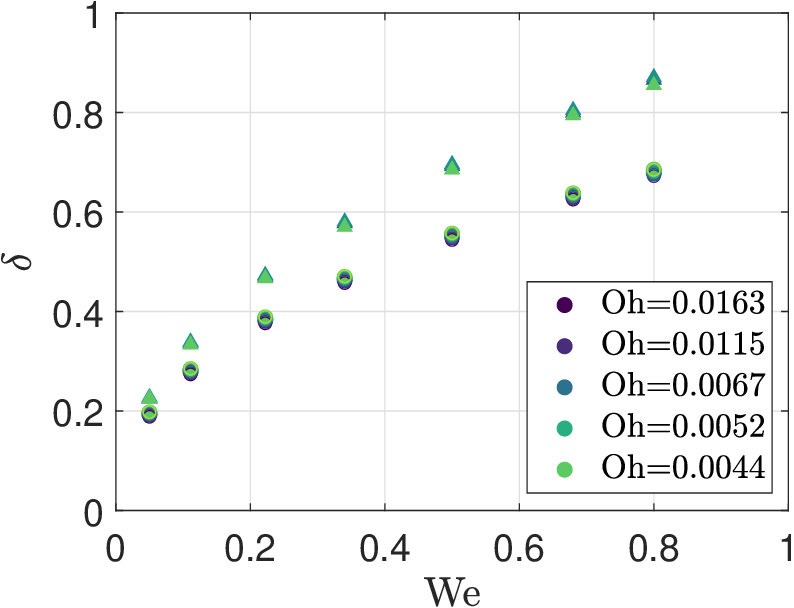}
        \end{subfigure}
        \begin{subfigure}{0.49\linewidth}
        \caption{}
            \includegraphics[width = \textwidth]{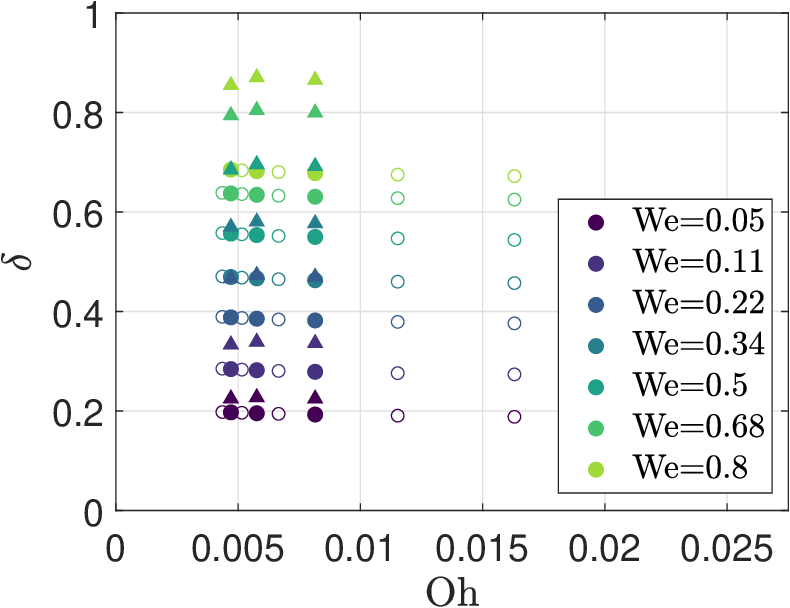}
        \end{subfigure} 
        \begin{subfigure}{0.49\linewidth}
        \caption{}
            \includegraphics[width = \textwidth]{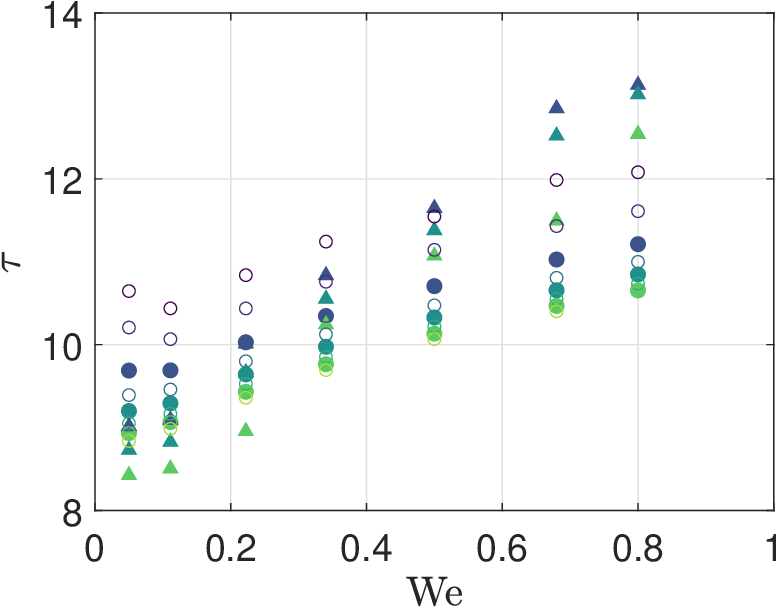}
        \end{subfigure}
        \begin{subfigure}{0.49\linewidth}
        \caption{}
            \includegraphics[width = \textwidth]{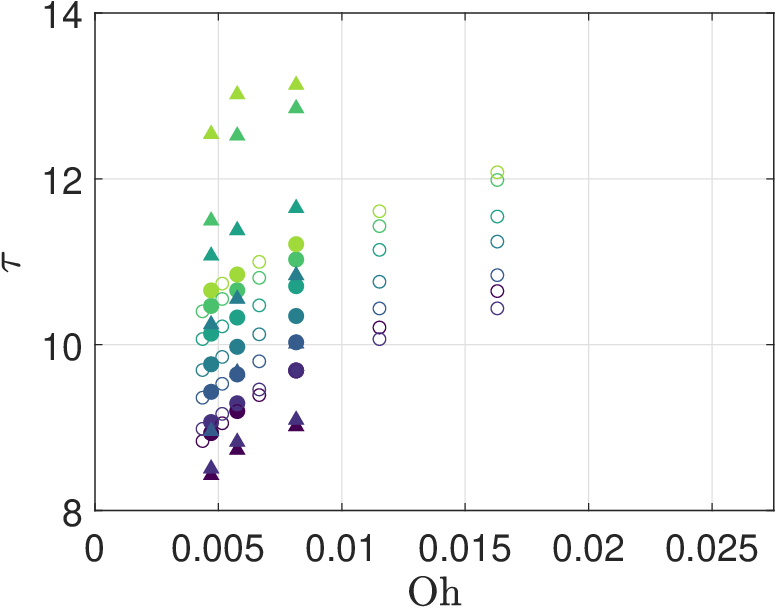}
        \end{subfigure}
        \caption{Comparisons of Coefficient of restitution $\alpha$, penetration depth $\delta$ and dimensionless contact time $\tau$, to Weber number (a,c,e) and Ohnesorge number (b,d,e) for a parameter sweep of values. The dimensionless parameter not plotted is indicated by the colour gradient on the opposite graph. Lubrication-mediated numerical results are presented by circles ($\bullet$), and triangles are used to indicate DNS results ($\triangle$) which correspond to a subset of the deformable parameter cases. The respective model subset is indicated by the circles being filled.}\label{fig: Comp We Oh}
    \end{center}
\end{figure}

\begin{figure}
    \begin{center}
        \begin{subfigure}{\linewidth}
            \includegraphics[width = \textwidth]{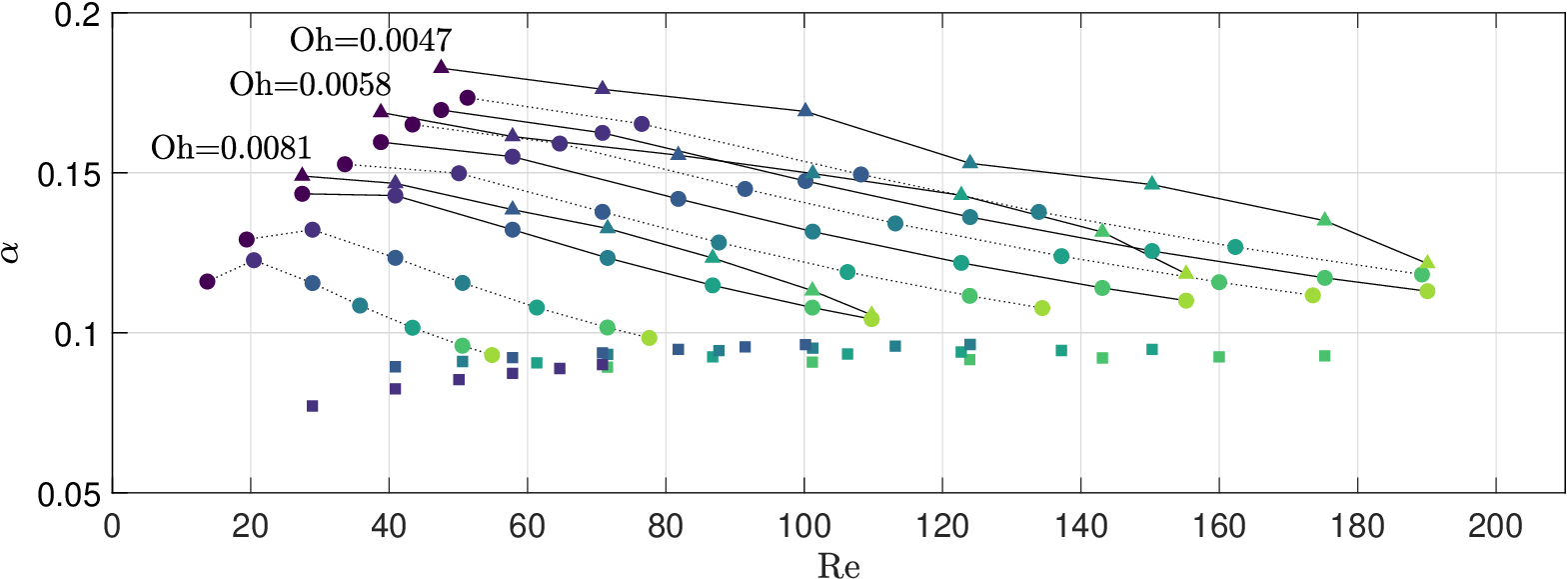}
        \end{subfigure}
        \begin{subfigure}{\linewidth}
            \includegraphics[width = \textwidth]{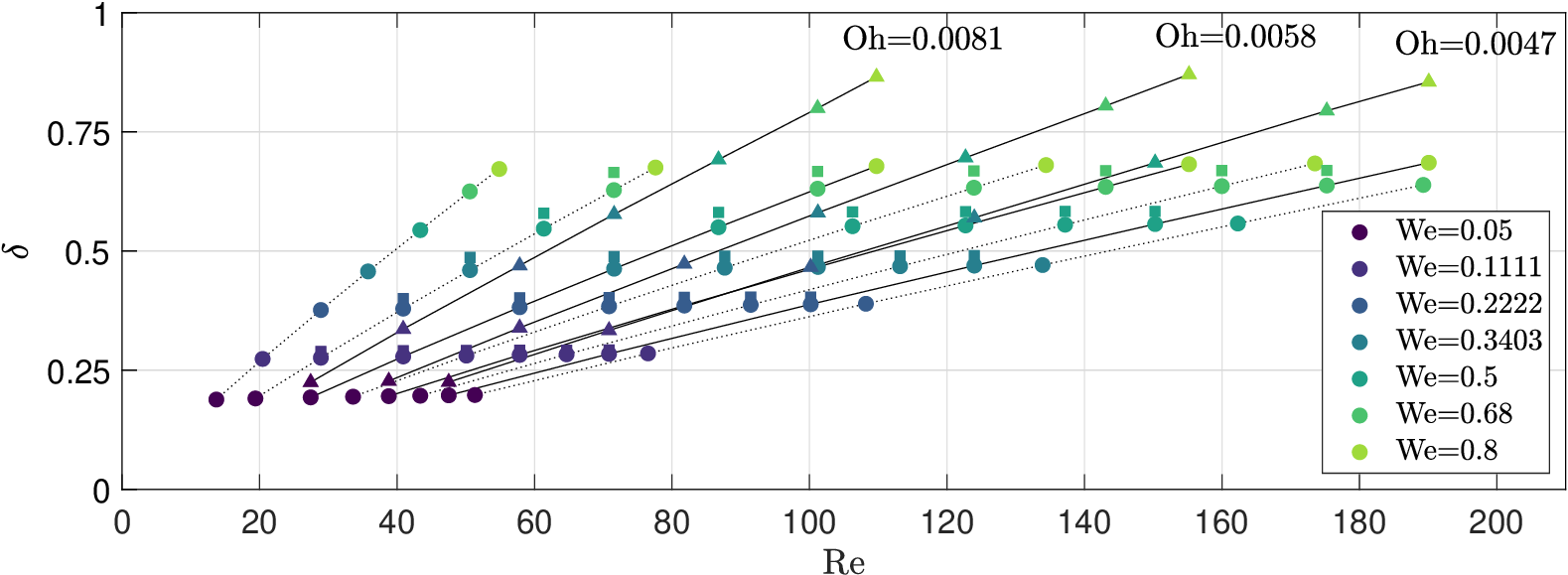}
        \end{subfigure}
        \begin{subfigure}{\linewidth}
            \includegraphics[width = \textwidth]{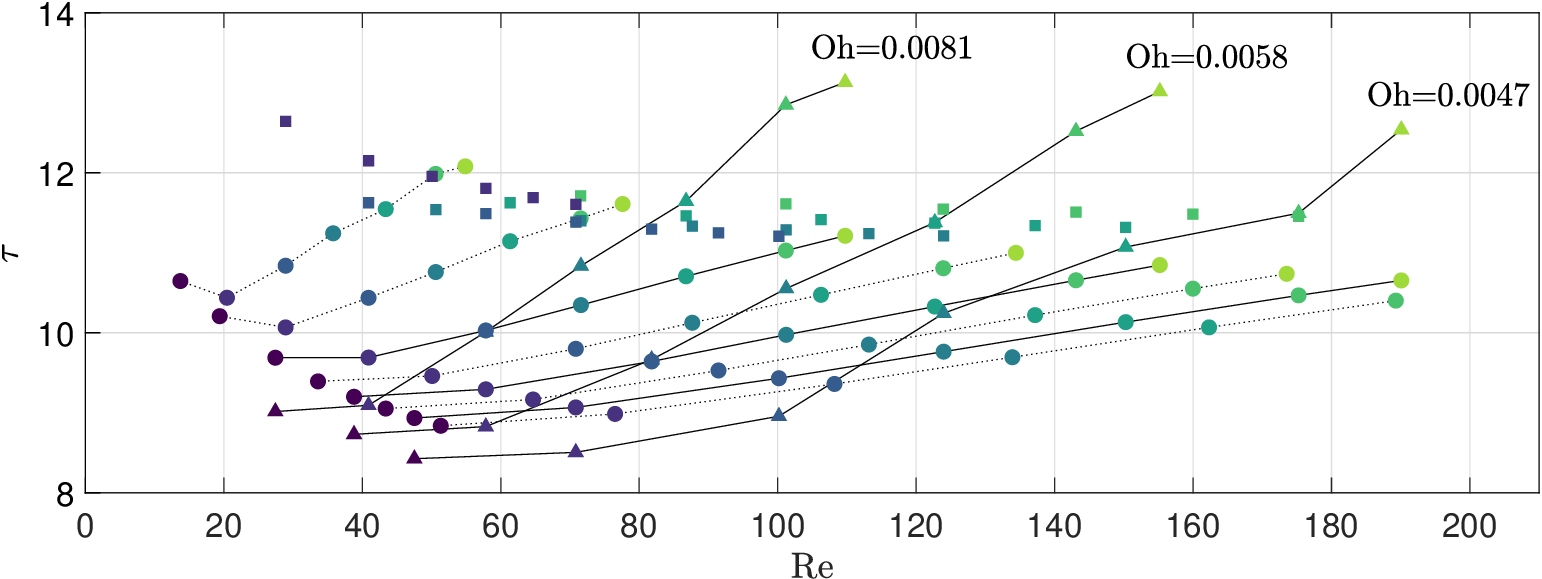}
        \end{subfigure}
        \caption{Coefficient of restitution $\alpha$, penetration depth $\delta$ and dimensionless contact time $\tau$ plotted against Reynolds number $Re$. The colours correspond to varying Weber numbers $W\!e$, with connected lines corresponding to fixed Ohnesorge numbers $Oh$. Here we have included the solid model results by the squares ($\square$), and as before circles ($\bullet$) denote the deformable model results, with solid lines highlighting their DNS counterpart results shown with triangles ($\blacktriangle$).}\label{fig: Reynolds}
    \end{center}
\end{figure}

Figures~\ref{fig: Comp We Oh} and \ref{fig: Reynolds} display the three parameters of interest $\alpha$, $\delta$, $\tau$, against $W\!e$, $Oh$, $Re$. We include solid and drop simulations for the lubrication-mediated model, and drop simulations for DNS.

From Figure~\ref{fig: Rebound Comparison (t,z)} we observed that
for a fixed parameter choice the solid sphere rebounded less (had a lower coefficient of restitution) than its deformable counterpart. This is demonstrated again within Figure~\ref{fig: Reynolds}, where solid rebounds are characterised by smaller coefficient of restitution values. We note that within the study the solid rebound cases present little variance in $\alpha$, except for a dropoff at lower $Re$. 

For the  deformable droplet, as shown in Figure~\ref{fig: Comp We Oh}, the coefficient of restitution generally decreases with $Oh$ and $W\!e$ (with the other parameter fixed). Exception to this are the larger Ohnesorge cases (corresponding to smaller droplets) where there is a local maximum in $\alpha$ as a function of $W\!e$. This has been observed in DNS and kinematic match modelling of three-dimensional droplets \cite{alventosa2023inertio}. In the respective work  axisymmetric 3D simulations (without resolving a lubrication layer by using a kinematic match model) are compared to other studies, including experimental observations. Broadly, between the three-dimensional data therein and our two-dimensional computations, general trends are preserved despite the restitution coefficients in three dimensions being approximately a factor of two larger than in two dimensions. Our lubrication-mediated two-dimensional model results and trends corroborate far better with the two-dimensional DNS results herein, the only differences being that DNS results are monotonic in $W\!e$ and have slightly higher coefficients of restitution, particularly at lowest $Oh$ and intermediate $W\!e$. 

 Maximum depth results, as shown in Figure~\ref{fig: Comp We Oh}, indicate strong agreement on trends between the model and DNS. Although the lubrication-mediated model always reached a lower depth than the DNS results, particularly at higher $W\!e$, both models show that penetration depth is nearly independent of $Oh$ and has a strong positive correlation with $W\!e$. We note that a similar discrepancy between DNS and the kinematic match model was observed in Alventosa \textit{et al} \cite{alventosa2023inertio} and that the discrepancy fell within the experimental range. As shown in Figure~\ref{fig: Reynolds}, the difference in maximum depth between the solid and deformable lubrication-mediated results was small.

The contact time $\tau$ varies slowly with $W\!e$ and $Oh$ numbers, showing a general slow increase with both. The DNS and model results are consistent with each other, with only minor differences. In the small $W\!e$ limit, we observe a local minimum in $\tau$ for the larger $Oh$ values, correlating with the maximum of $\alpha$. There is an inverse relationship between contact time and coefficient of restitution: indeed a good portion of the contact time as defined here is post detachment when the droplet is still below the undisturbed free surface. As seen from Figure~\ref{fig: Droplet Deformations} in that case the actual impact takes about $5 t_\sigma$. The contact time is even longer for the solid impacts, as the impactors are taking even longer to reach this $z=R_0$ threshold criteria. 

Overall the agreement between the LM model and the DNS is highly encouraging, with key trends replicated successfully and explainable discrepancies owing to modelling assumptions or anticipated numerical effects on both sides. More specifically, DNS results can be summarised as exhibiting stronger pool deformation on impact, larger coefficients of restitution and a wider variation interval in contact times than their reduced-order model counterparts. Figure~\ref{fig:varyingviscosity} in Appendix~\ref{app: Visc} encapsulates some of the above features for a typical comparison case. The viscosity enhancement in the model provides a useful handle for navigating the deformable to non-deformable impactor property set, while addressing numerical challenges in terms of stability in its discretisation scheme. The lower the value of this parameter (down to a tractable threshold level), the closer the results become to the DNS benchmark liquid impact calculation, with pool deformation depth converging towards DNS-obtained results for the lowest enhanced viscosity values used in the model in the liquid impact case. The scenarios described by more pronounced pool deformation naturally lead to stronger bouncing, expressed in terms of larger coefficients of restitution, with any differences arguably becoming more prominent as $W\!e$ increases and fully nonlinear DNS calculations are expected to start diverging from the simplified model counterpart results. The behaviour of the impactor during the spreading and rebound stages in the two approaches is another point of differentiation, with more nonlinear behaviour in moderate to large deformation cases in the DNS datasets visible even as part of earlier air layer height dynamics studies, as shown for example in Figure~\ref{fig: Air Layer (x,t)} in terms of the radial extent of the contact region. More deformed droplets on impact in the DNS lead to larger contact radii and marginally thinner air layers, while the rebound phase is also slightly accelerated by comparison, correlating with the earlier observations on the small coefficient of restitution differences between the two employed methodologies.

\begin{figure}
    \begin{center}
        \begin{subfigure}{0.49\linewidth}
            \caption{}
            \includegraphics[width = \textwidth]{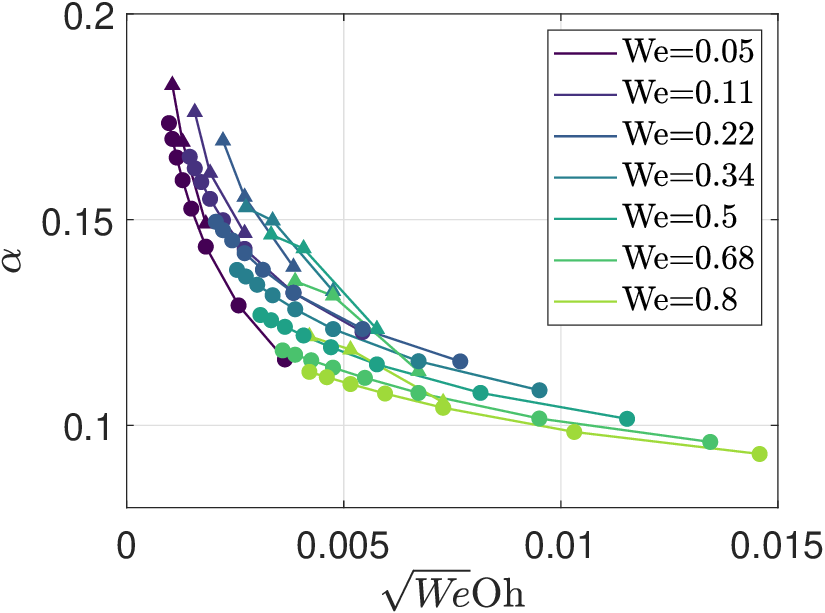}
        \end{subfigure}
        \begin{subfigure}{0.49\linewidth}
        \caption{}
            \includegraphics[width = \textwidth]{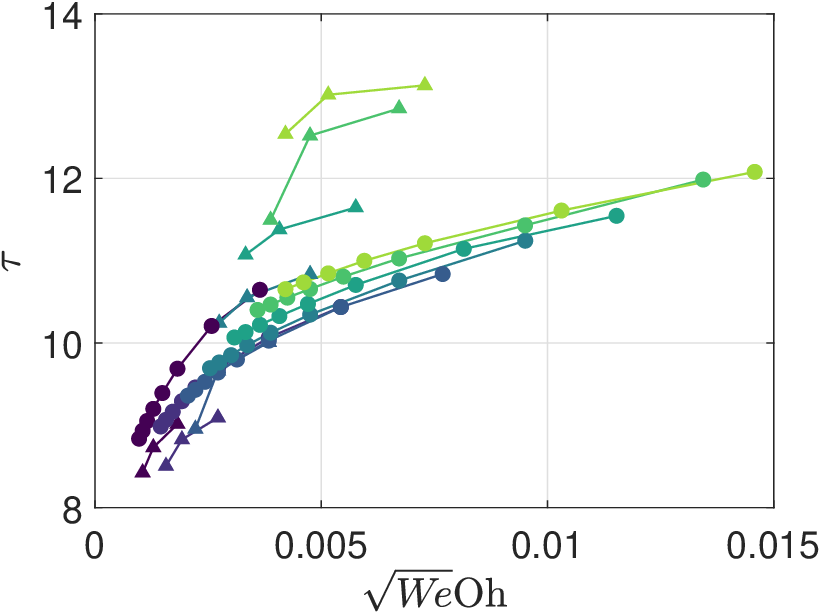}
        \end{subfigure}
        \caption{Collapsed data for coefficient of restitution $\alpha$ and dimensionless contact time $\tau$, both plotted against $\sqrt{W\!e}\,Oh$. The colour gradient corresponds to Weber number $W\!e$. The circles $(\bullet)$ denote deformable LM model results, and DNS data is shown using triangles $(\blacktriangle)$.}\label{fig: Collapsed}
    \end{center}
\end{figure}

We  postulate that the impact depth $\delta$ depends little on $Oh$ because the maximum depth is achieved early in the impact cycle. Thus most energy loss up to this stage is due to the transfer of energy to motion in the bath and not to viscous effects. One might expect $\delta \sim \sqrt{W\!e}$ from a model of a linear spring whose initial condition is proportional to the impact velocity $W_0$. The contact time and coefficient of restitution, however, depend more strongly on viscosity. When the data is plotted against $\sqrt{W\!e} \, Oh$ in Figure~\ref{fig: Collapsed} we observe a better description of the behaviour and collapse of the data. A justification for this scaling arises from postulating an exponential  decay rate for a linear spring model with decay rate proportional to $Oh$. This results in a linearly approximated viscous loss proportional to $Oh$, for small $Oh$ and finite time. For the contact times, we note that some high Weber number DNS data fall above lower Weber number data in this scaling. We conjecture that this is due to the increase in the (integer) number of oscillations of the droplet during impact in the DNS. These additional oscillations are not present in the LM model due to the additional damping effects introduced there.

\section{Repeated bouncing}\label{sect:Faraday}

In the previous section we presented a series of results obtainable by the lubrication mediated model and highlighted their general agreement with well validated DNS methods through a parameter study. We will now highlight the robustness of the lubrication mediated model through demonstrating sustained bouncing behaviour under the influence of gravity and a vertical oscillation of the bath. This justifies the potential of the model to be used within further studies of Faraday pilot-wave models: in particular tunnelling studies \cite{nachbin2017tunneling} use two-dimensional droplets. 

Thus far we have reviewed the results of the lubrication-mediated model of the two-dimensional droplet rebounding off a deep liquid bath. A natural extension to this system is the inclusion of repeated bounces with the inclusion of other effects. With the addition of gravity, after the initial rebound and detachment, the droplet undergoes ballistic motion and falls back onto the bath. As noted by Couder \textit{et al} \cite{couder2005walking}, a way to investigate repeated bouncing behaviour of droplets is to vibrate the liquid bath and achieve a sustained periodic behaviour for a vibrational forcing of sufficient amplitude. Both gravity and vibrational forcing are straightforward additions to the model and a sinusoidal forcing in the gravity term is introduced with the functional form $G(t) = g(1 - \Gamma \cos(\omega_f t))$.  For sufficiently large $\Gamma$ such a forcing generates the Faraday instability: growing standing waves on the free surface known as Faraday waves. The threshold at which Faraday waves are excited is denoted $\Gamma_F$. In order for sustained bouncing behaviour to occur, the bath is instead oscillated close to, but below this threshold, that is, for some $\Gamma < \Gamma_F$. In what follows we present the results when such a forcing is introduced to the lubrication-mediated model. 

\begin{figure}
    \centering
    \includegraphics[width=\textwidth]{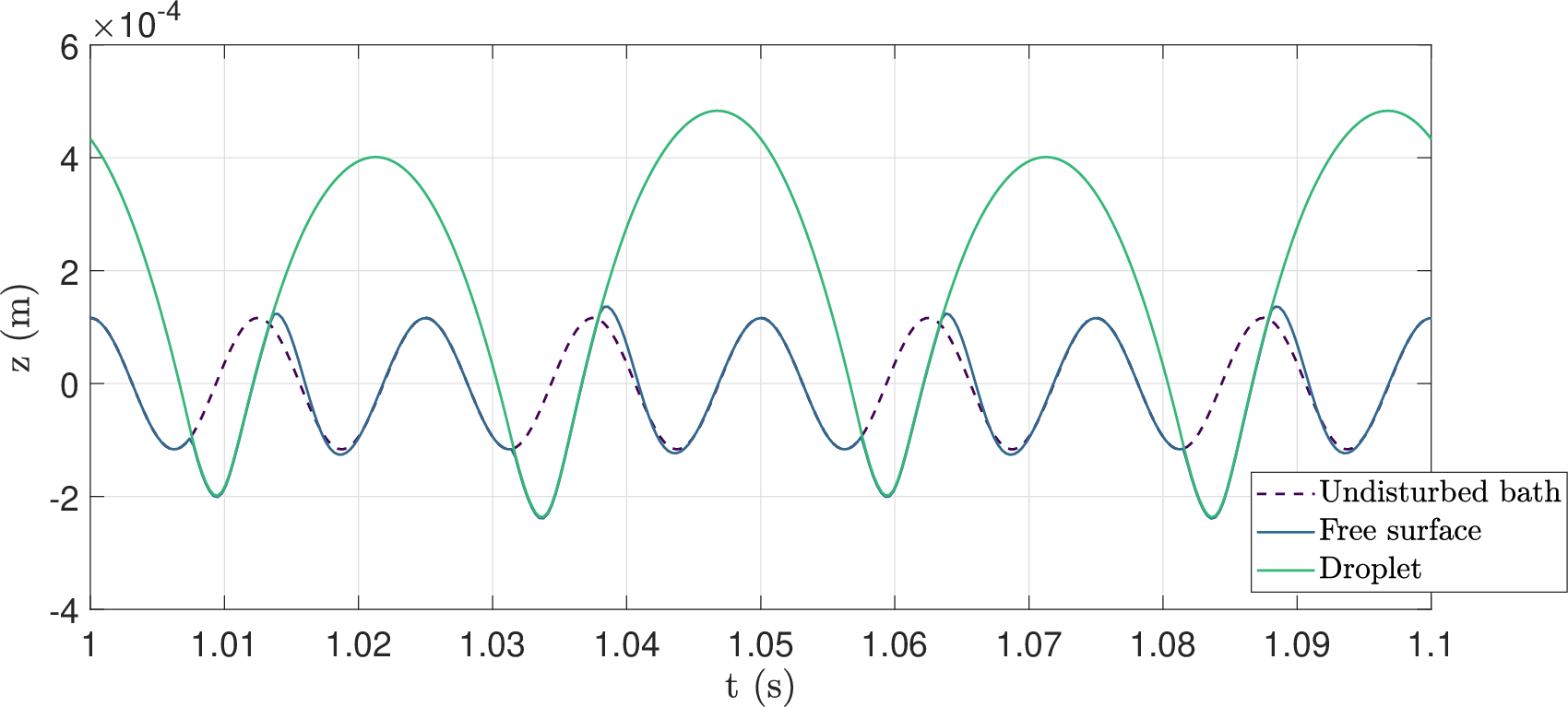}
    \caption{Periodic droplet rebounds obtained for viscous silicone oil with radius $R=0.83$mm, on a bath oscillated at $80$Hz with forcing amplitude $\Gamma = 3$. Plotted with the oscillation period of an undisturbed free surface behind to highlight deviation away from this rest state. The dynamics settles on a periodic pattern corresponding to 4 oscillations of the bath every 2 rebounds of the droplet (denoted as a (4,2) mode), within a period $0.05$ s.}
    \label{fig: faraday}
\end{figure}

Following the methodology presented in Milewski \textit{et al} \cite{milewski2015faraday}, we incorporate the forced oscillations of the bath as an oscillating frame of reference. For a given oscillation frequency $\omega_f$ and forcing amplitude $\Gamma$, the equations derived in Section~\ref{sect:Model} for the liquid bath are now given as 
\begin{align}
\partial_t \eta_b  &= \partial_z \phi + 2\nu_b \partial_x^2 \eta_b, \qquad &z&=0, \label{eq: Faraday KBC}\\
\partial_t \phi &= -\frac{1}{\rho_b} p_a -  g(1-\Gamma \cos(\omega_F t))\eta_b + \frac{\sigma_b}{\rho_b}\partial_x^2 \eta_b  + 2\nu_b \partial_x^2 \phi, \qquad &z&=0, \label{eq: Faraday  DBC}
\end{align}
and the equation governing the droplet motion $\tilde{Z}(t)$ in the oscillating frame are
\begin{equation}\label{eq: Faraday vert dyn}
    \rho_d \pi R_0^2 \frac{\text{d}^2 \tilde{Z}}{\text{d}t^2} = \int_{-l^*}^{l^*} P dx - \rho_d \pi R_0^2 g(1-\cos(\omega_F t)).
\end{equation}
The motion in a fixed laboratory frame $Z(t)$ is obtained by the transformation $Z = \tilde{Z} + \Gamma \omega_f^{-2} \cos(\omega_f t)$. Figure~\ref{fig: faraday} demonstrates the periodic bouncing behaviour obtained with the lubrication-mediated model. To capture these results, the numerical domain was shortened to an integer ratio of the Faraday wavelength of the bath, specifically, corresponding to $L = 12$cm. The droplet was then released from a small height, with an initial velocity of $0.1$ms$^{-1}$. The choice of initial velocity is unimportant as long as the droplet is able to rebound, as the system will eventually stabilise to a periodic state with these parameters. 

As shown in Figure~\ref{fig: faraday} the stable state is a $(4,2)$ mode \cite{milewski2015faraday}, which is denoted as such due to the periodic motion requiring 4 oscillations of the bath for every 2 rebounds of the droplet. In the figure, the dashed line indicates the unperturbed surface, as though the behaviour of the droplet were omitted from the system. The upper line then represents the south pole of the drop, and the lower blue line corresponds to the free surface of the liquid bath.

We note that the full simulation was run for $1.2$ seconds of physical time in order for the system to reach periodicity from an arbitrary initial data. Thus this simulation was between 10-100 times longer in physical time than the single-bounce simulations in our study. Despite this, the simulation only required several hours of runtime on a standard desktop in MATLAB. Such a simulation would be prohibitively expensive for DNS. 

\begin{figure}
    \begin{center}
        \begin{subfigure}{0.425\linewidth}
          \subcaption{}
            \includegraphics[width=\textwidth]{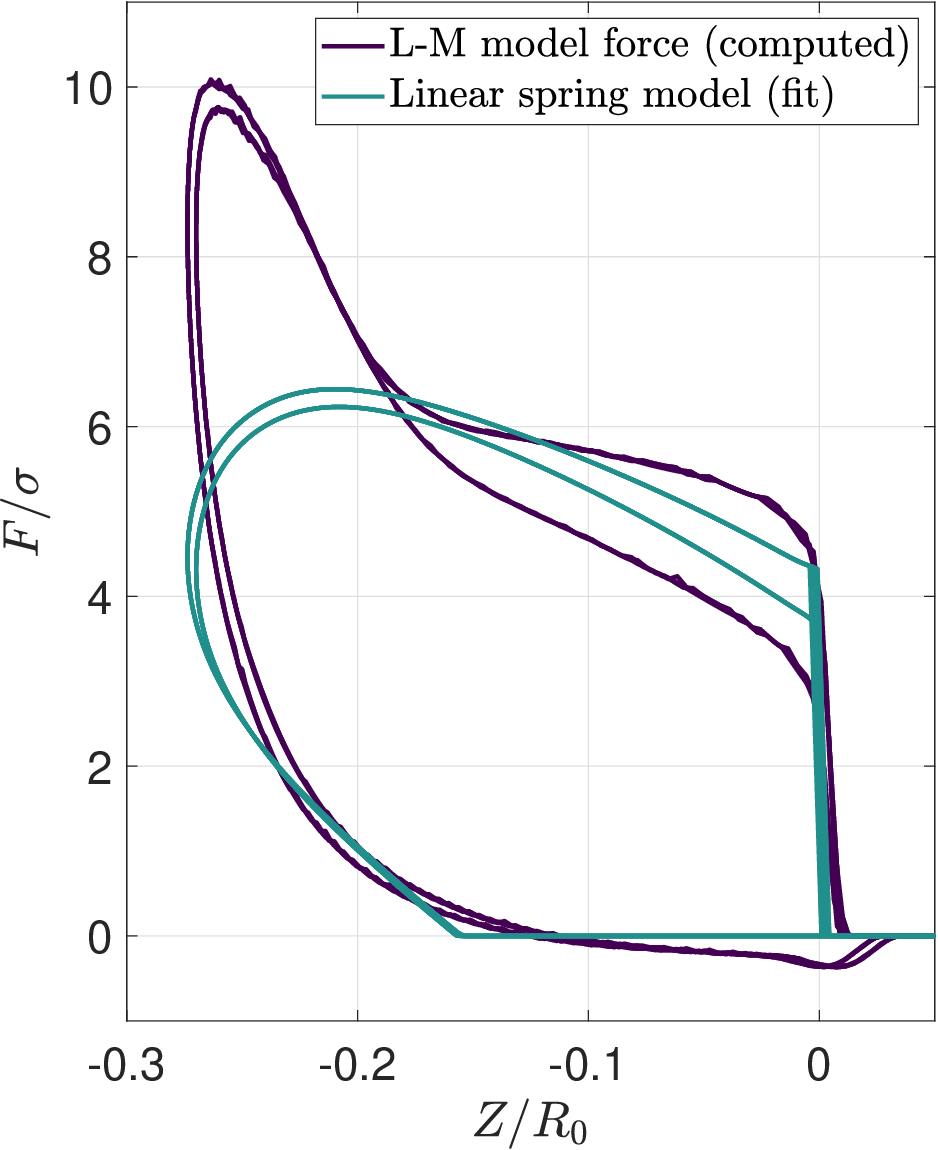}
        \end{subfigure}
        \begin{subfigure}{0.565\linewidth}
            \subcaption{}
            \includegraphics[width=\textwidth]{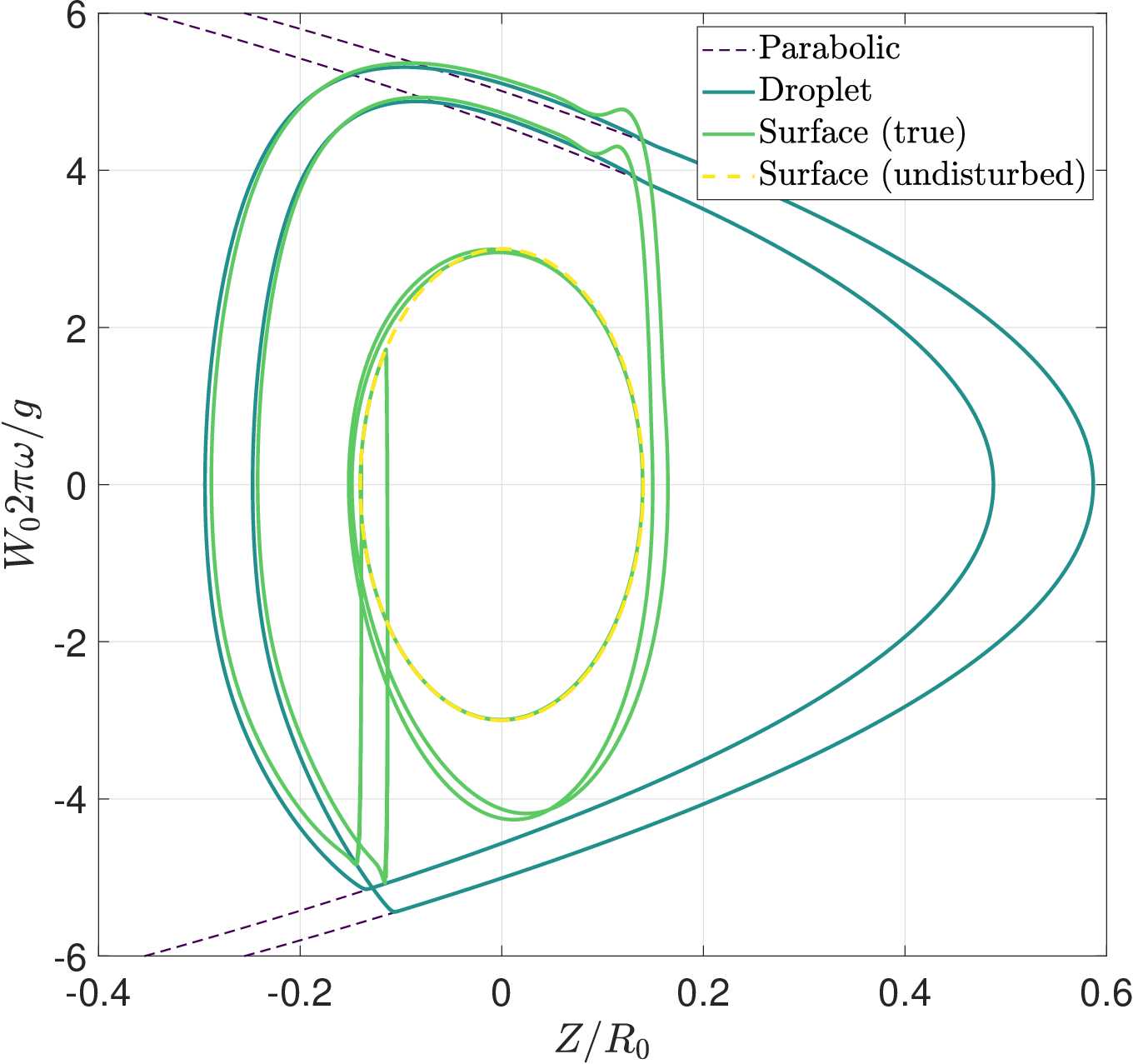}
        \end{subfigure}
    
    \caption{(a) Computed Force-depth plot for L-M model in a (4,2) mode (hence 2 cycles in the figure per period of motion) and fitted data for a linear spring model. (b) Phase-plane dynamics in an inertial frame of reference for the drop and the free-surface. 2 cycles of the (4,2) mode are shown. The drop alternates from a parabolic trajectory (shown together with a dashed line prediction) and an impact state. The free surface alternates between an elliptic trajectory (prediction shown in dashed ellipse in the centre of the figure) and being almost equal to the drop during impact.}
    \label{fig: Force Plot}
    \end{center}
\end{figure}

In several studies of bouncing drops, a linear spring model has been used to elucidate the effect of impacts, for example off soap films \cite{gilet2009fluid} or to describe drops rebounding on a solid hydrophobic surface \cite{okumura2003water}. A linear model for rebounding off vibrating baths was proposed in the work of  Mol{\'a}{\v{c}}ek et al. \cite{molavcek2013dropsA}, and recently Primkulov et al. \cite{primkulov2025nonresonant} used such a model to describe non-resonant effects in pilot wave hydrodynamics. The simplest linear spring model takes the form
\begin{equation}
m\frac{d^2 Z}{dt^2} + \gamma \frac{d\bar{Z}}{dt} + \beta \bar{Z}   = -mg, \qquad \bar{Z}<0.
\end{equation}
Here $Z$ is measured in an inertial frame, while $\bar{Z}$ and $\frac{d\bar{Z}}{dt}$ are measured relative to the free surface; $\bar{Z}$ is the depth of impact, and $\frac{d\bar{Z}}{dt}$ is the speed of the drop relative to the free surface speed. These quantities are somewhat ambiguous since the impact itself deforms the surface, but may be approximated as relative to the free surface \emph{unperturbed} by the impact. The forces exerted by the bath on the droplet are therefore given by 
$$- \left(\gamma \frac{d\bar{Z}}{dt} + \beta \bar{Z}\right), \qquad \bar{Z}<0. $$

From our numerical simulations of the LM model, we recover the force $F(t) = \int_{-l^*}^{l^*} P dx$ experienced by the droplet, its impact depth $\bar{Z}(t)$ and velocity $\frac{d\bar{Z}}{dt}$ and attempt to construct a plausible linear model by fitting $\gamma$ and $\beta$. 

The results are presented in Figure~\ref{fig: Force Plot}, where we show the computed force during impact in our LM model for the simulation in Figure~\ref{fig: faraday}. The periodic curves in the $\bar{Z}-F$ plane are composed of two cycles, one for each impact. For each cycle there is a jump in the force at impact, and this elevated force persists (and increases) throughout the downward motion of the drop. During upward motion the force decreases to a much lower level. The area inside each cycle is the net work done on the fluid bath by the droplet, generating the wave motion. An approximate linear fit of the force curve may be made using $\gamma \approx 0.35$ and $\beta \approx 20$ as shown in the figure. While there appear to be some nonlinear effects at play in the dynamics shown, it is plausible that a linear model could provide further simplifications and efficiencies while also capturing essential behaviour. Implementing a linear model including wave generation is beyond the scope of this work. 

Figure~\ref{fig: Force Plot} also shows the corresponding trajectory of our computed LM solution in the inertial $Z-W$ phase plane. The droplet exhibits parabolic flight trajectories and a clear jump in its acceleration at impact, while the free surface has a jump in velocity at the same time. The dashed lines in the figure indicate parabolas matching the flight path of the droplet and an elliptical path marking an undisturbed vertically oscillating free surface (both in a laboratory frame). The figure demonstrates that the free surface returns to this undisturbed oscillation rapidly after droplet lift-off and before the next impact.
\section{Conclusion}\label{sect:Conclusion}

In this theoretical investigation we have developed a new approach for considering the air layer trapped between an impacting drop and a liquid pool, which permits the study of rebound off a deep liquid bath in a more physically complete manner relative to methodologies which ignore the presence of the gas altogether. Through evolving each of the three regions separately and providing a method of coupling through the pressure obtained from a thin film approximation for the air, we have shown that it is possible to obtain good agreement with results obtained via direct numerical simulation.

The lubrication-mediated model that we have introduced has the capability of capturing the height and pressure within the air layer through the solution of an elliptic thin film problem over a dynamic boundary, which is novel in this context. We are able to solve for solid impactors and for deformable droplets in the small deformation limit. Our model is also restricted to small bath deformations and small viscous effects.

We have shown through an in-depth comparison for a single rebound case that the lubrication-mediated model is able to qualitatively match with the DNS system, obtaining similar behaviours and dynamics. A highlight is the comparison of air layer thicknesses shown in the previous section, which has good agreement with the morphology of air layers seen in droplet rebound experimental \cite{tang2019bouncing} and DNS studies. 

 Across a range of parameters we have uncovered that the lubrication-mediated model is able to capture behaviour of models based on the kinematic match approach, matching qualitatively with the respective 3D reduced methods \cite{galeano2021capillary, alventosa2023inertio}, yet providing additional insight into the air layer dynamics without increased computational effort. 
 We have discussed the numerical stiffness observed in our system, and have proposed viscous damping methods to mitigate this issue. Further investigation is needed to improve the method and adapt it to regimes which permit larger deformation of the drop and liquid bath. 

 There are several avenues to extend this work, and the related 3D rigid-sphere axisymmetric computations in \cite{phillips2024lubrication}. Two direct extensions are three-dimensional droplet impacts and two-dimensional non-symmetric impacts.

 Finally, we have demonstrated the robustness of the lubrication-mediated model, and its applicability to repeated bouncing behaviour under sustained forcing as first discovered  by Couder \textit{et al} \cite{couder2005walking}. Future avenues of work  would be to investigate the results of the lubrication mediated model in Faraday pilot wave studies such as for walkers where almost vertical impacts capture the dynamics \cite{galeano2017non}, and in dynamics where the vertical motion is known to play an important role such as in corrals \cite{durey2020faraday}, tunnelling \cite{nachbin2017tunneling} and non-resonant behaviour \cite{primkulov2025nonresonant}. 


\appendix
\section{Physical parameters}\label{app: Table}
Table \ref{tab: Appendix} lists the parameters used in the present study.
\begin{table}
\begin{center}
\begin{tabular}{l l c l}
\hline
Parameter  & Symbol &  Value & Units  \\
\hline
\hline
Density (air) &  $\rho_a $  & 1.225 & kg m$^{-3}$\\
Dynamic viscosity (air) & $\mu_a $  & $1.825\times 10^{-5}$ & kg m$^{-1}$s$^{-1}$\\

Density (water) &   $\rho_\beta $  &1000 &kg m$^{-3}$\\
Dynamic viscosity (water)  &   $\mu_\beta $  & $8.94\times 10^{-4}$& kg m$^{-1}$s$^{-1}$\\
Density (silicone) &   $\rho_\beta $  &949 &kg m$^{-3}$\\
Dynamic viscosity (silicone) &  $\mu_\beta $  & $1.9\times10^{-2}$ &kg m$^{-1}$s$^{-1}$\\ 

Surface tension (air-water)  &  $\sigma_\beta $  & $7.49\times10^{-2}$ &kg s$^{-2}$\\
Surface tension (air-silicone) &  $\sigma_\beta $  & $2.06\times10^{-2}$ &kg s$^{-2}$ \\
\hline
Drop radius (water) &  $R_0 $  & $0.5\times 10^{-4}$ - $7\times 10^{-4}$ &m\\
Initial speed (water) &   $W_0 $  & 0.0717 - 1.0733 &m s$^{-1}$\\
Weber Number (water) &  We  & 0.05 - 0.8 & --\\
Ohnesorge Number (water) &  Oh  & 0.0044 - 0.0163 &--\\
Reynolds Number (water) & Re & 13.7 - 190.1 &--\\
 \hline
 Drop radius (silicone) &  $R_0 $  & $8.3\times 10^{-4}$ &m\\
Initial speed (silicone) &   $W_0 $  & 0.2-0.4 &m s$^{-1}$\\
Weber Number (silicone) &  We  &  1.54 - 6.11 & -- \\
Ohnesorge Number (silicone) &  Oh  & 0.149 & --\\
Reynolds Number (silicone) & Re & 1.53 - 16.583 & --\\
\hline
\end{tabular}
\caption{Physical parameter values for the results presented within this paper, including value ranges for nondimensional numbers for the chosen parameters. The values of water were used throughout the main body of the text in section \ref{sect:Results}, with the switch to silicone oil used in the extension to Faraday pilot-waves in section \ref{sect:Conclusion}. Silicone oil is used experimentally and, due to the increased viscosity and decreased surface tension, also results in less stringent numerical timestepping. }\label{tab: Appendix}
\end{center}
\end{table}

\section{Boundary layer scaling}\label{App: Boundary}
Solutions to the quasi-potential model laid out by equations \eqref{eq: laplace}-\eqref{eq: end of full system} exhibit a boundary layer effect at higher Reynolds numbers. In order to see that one may solve the (linear) `free-wave' problem and obtain explicitly the forms of the free surface $\eta_b$ and velocity potential $\phi_b$ as 
\begin{equation*}
    \eta_b = a e^{i(kx \pm \omega_0 t)} e^{-2\nu_b k^2 t}, \qquad \phi_b = \pm i a \frac{\omega_0}{|k|} e^{i(kx \pm \omega_0 t)} e^{-2\nu_b k^2 t} e^{|k|z},
\end{equation*}
where $\omega_0 = \sqrt{|k|(g + \sigma_b k^2/\rho_b)}$ is the frequency of undamped capillary-gravity waves. As expected, this is a time decaying travelling wave. Furthermore, one may now compute the induced boundary layer represented by $\psi(x,z,t)$ using $\partial_x \psi(x,0,t) = 2\nu_b \partial_x^2 \eta_b(x,t)$ (from the leading order balance of (2.17)). This results in 
$$\Psi = 2ik\nu_b a e^{i(kx \pm \omega_0 t)} e^{-2\nu_b k^2 t}e^{\pm i\sqrt{2}\sqrt{Re_\omega}|k|z/2}e^{ \sqrt{2}\sqrt{Re_\omega}|k|z/2},$$
where $Re_\omega = \frac{\omega_0}{\nu_b k^2}$. Hence one finds that the vertical decay length scale for $\psi$ is $O(Re_\omega^{-1/2})$ and that denoting $u^\psi,v^\psi$ and $u^\phi,v^\phi$ as the vortical and irrotational components of the velocity field respectively, we have
$$u^\psi \sim Re_\omega^{-1/2} u^\phi, \qquad v^\psi \sim Re_\omega^{-1} v^\phi.$$
This describes the boundary layer structure as $Re_\omega \rightarrow \infty$.

\section{Viscosity consideration}\label{app: Visc}
 
The most challenging numerical issue we encountered was an instability related to stiffness of this problem. The severity of this issue depended on $W\!e$ and $Oh$, mainly for high $W\!e$. Instabilities correlated with larger deformations of the droplet which, through the thinning of the lubrication layers creates large pressure fluctuations from the $h^{-3}$ dependence of pressure on layer thickness. Naturally, this issue was absent in the solid impact case. 

We chose to resolve this issue with an artificially enhanced viscosity during contact, specifically while $h \le \varepsilon R_0$, and describe here the choice of viscosity and its effect on the results of a test deformable case. An in-depth stability analysis of the system and other solutions for this numerical issue is left for future work. Briefly, one could consider fully implicit numerical schemes (which pose some challenges due to the nonlinearity of the elliptic lubrication equation), split step schemes where the drop oscillations and lubrication equation are resolved implicitly or further asymptotic reduction of the droplet equations during impact.

\begin{figure}
    \centering
    \includegraphics[width=0.9\textwidth]{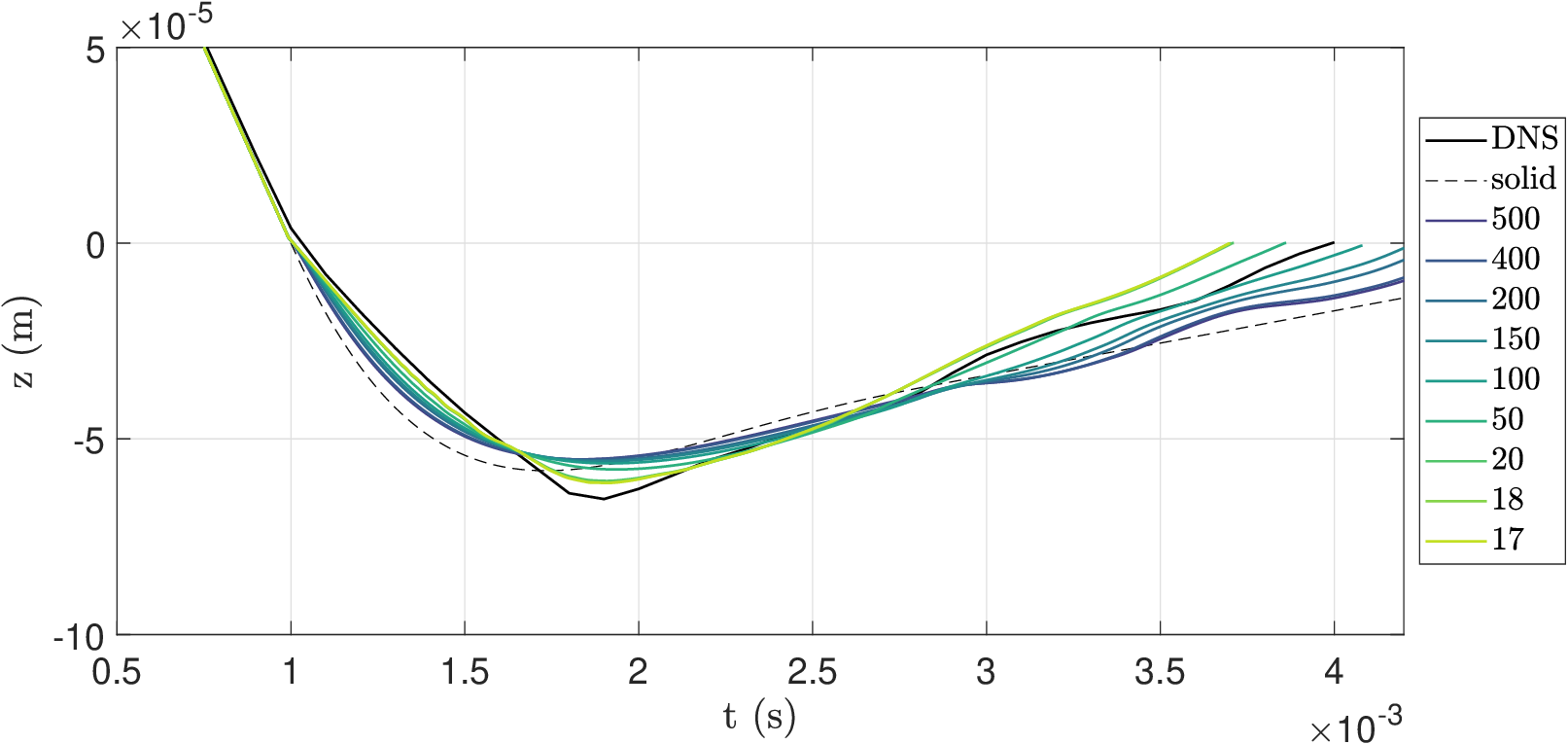}
    \caption{Trajectories for droplets with $R_0= 0.2$mm, and $W_0 = -0.2$ms$^{-1}$, using a modified viscosity $\mu^* = A\mu$, with $A$ ranging from $17$ to $500$. The solid black line corresponds to DNS results, and the dashed black line corresponding to the solid rebound of the same size and initial velocity.}
    \label{fig:varyingviscosity}
\end{figure}

We introduce a viscosity enhancement coefficient $A$,  such that $\mu^* = A\mu$ during contact. We determined the smallest value of $A$ for which the entire parameter sweep of simulations would be stable, through over-damping the droplet and slowly lowering the value of $A$. Figure~\ref{fig:varyingviscosity} demonstrates a range of values of $A$ for the single run case with $R_0 = 0.2$mm and $W_0 = -0.2$ms$^{-1}$. We present the solid lubrication mediated model with corresponding values, alongside DNS of the same initial conditions. The solid model can be interpreted as the limit $A \to \infty$. For this case, the choice $A=17$ was the lowest value that could be used reliably without introducing numerical instabilities. By contrast, large values of $A>400$ seemed to converge to the same result. Considering the full $W\!e$ and $Oh$ parameter regime within our study, the choice $A=150$ provided the smallest damping without instability. 

In Figure~\ref{fig:varyingviscosity} we observe the effect that the damping has on the rebound, specifically plotting the vertical height of the south pole. The DNS result which does not have a modified viscosity has a pronounced and sharper deflection at first point of contact of the drop and bath. By contrast, the solid drop case leads to a smooth trajectory.  The switching off of the additional viscosity post-contact allows for the oscillations that are expected, and are closer aligned to the behaviour of the DNS results. Further investigation of energy transfers during impact could lead to improved reduced models.

\paragraph{Funding:}K.A.P is supported by a scholarship from the EPSRC Centre for Doctoral Training in Statistical Applied Mathematics at Bath (SAMBa), under the project EP/S022945/1. R.C. gratefully acknowledges support from US-CBET/UK-EPSRC grant EP/W016036/1.


\bibliographystyle{unsrt}
\bibliography{bibs}

\end{document}